\documentclass[10pt]{article}
\usepackage{amsmath,amsfonts,xypic}
\usepackage{amssymb}
\usepackage{amscd}
\usepackage[all]{xy}
\advance \textheight by \topskip
\makeatletter
\@addtoreset{equation}{section}
 \def\theequation{\thesection.\arabic{equation}}
\makeatother

\newtheorem{theorem}{Theorem}[section]

\newtheorem{corollary}[theorem]{Corollary}

\newtheorem{definition}[theorem]{Definition}
\newtheorem{example}[theorem]{Example}

\newtheorem{lemma}[theorem]{Lemma}

\newtheorem{proposition}[theorem]{Proposition}
\newtheorem{remark}[theorem]{Remark}

\def\NN{{\mathbb N}}
\def\CC{{\mathbb C}}

\def\NN{{\mathbb N}}
\def\ZZ{{\mathbb Z}}
\def\ad{{\mathrm {ad~}   P}}
\def\ker{{\mathrm{Ker}}}
\def\K{{\mathbb K}}
\def\m{{\frak{g}}}
\newcommand{\beq}{\begin{equation}}
\newcommand{\eeq}{\end{equation}}
\newcommand{\beqa}{\begin{eqnarray}}
\newcommand{\eeqa}{\end{eqnarray}}
\newcommand{\noi}{\noindent}
\newcommand{\e}{\varepsilon}
\newcommand{\p}{\varphi}

\newcommand{\g}{{\mathfrak g}}

\newcommand{\pard}{[\!\! [}
\newcommand{\pari}{]\!\! ]}

\newcommand{\el}{elementary Lie algebra of order $3$ }

\newcommand{\s}{{{\mathfrak{sl}}(2)}}

\def\i{\frak{iso}(1,3,\mathbb C)}
\newcommand{\Sy}{{\cal S}}
\newcommand{\V}{{\cal V}}
\newcommand{\D}{{\cal D}}
\newcommand{\nse}{ {\subseteq \hskip -.3truecm \slash \ }}
\def\>{\rangle}
\def\<{\langle}

\begin{document}

\title{
{\bf  Poincar\'e and $sl(2)$ algebras of order $3$ }  }

\author{ 
{\sf   M. Goze} \thanks{e-mail:
m.goze@uha.fr}$\,\,$ ${}^{a}$,
{\sf  M. Rausch de Traubenberg }\thanks{e-mail:
rausch@lpt1.u-strasbg.fr}$\,\,$ ${}^{b}$
and
{\sf A. Tanas\u a}\thanks{e-mail:
adrian.tanasa@ens-lyon.org}$\,\,$ $^{a}$
\footnote{Michel Goze and Adrian Tanas\u{a} have been partially 
supported by the Inter-University 
Cooperation Program of AUF (Agence Universitaire de la Francophonie).} 
\\
{\small ${}^{a}${\it Laboratoire MIA,
Universit\'e de Haute Alsace,}} \\
{\small {\it Facult\'e des Sciences et Techniques,
4 rue des Fr\`eres Lumi\`ere, 68093 Mulhouse Cedex, France.}}  \\ 
{\small ${}^{b}${\it
Laboratoire de Physique Th\'eorique, CNRS UMR  7085,
Universit\'e Louis Pasteur}}\\
{\small {\it  3 rue de
l'Universit\'e, 67084 Strasbourg Cedex, France}}}
\date{\today}
\maketitle
\vskip-1.5cm

\vspace{2truecm}

\begin{abstract}
 In this paper we initiate a general classification for
Lie algebras of order 3 and we give all Lie algebras of order 3 
based on $\mathfrak{sl}(2,\mathbb C)$ and $\mathfrak{iso}(1,3)$ the 
 Poincar\'e algebra in four-dimensions.
We then
set the basis of the theory of the
deformations (in the Gerstenhaber sense) and contractions for Lie
algebras of order 3.
\end{abstract}

{\bf Pacs } 02.20Sv, 11.30.Ly 

{\bf keywords} Contractions, generalized In\"on\"u-Wigner, Lie algebras 
\maketitle

\section{Introduction and motivation}
\renewcommand{\theequation}{1.\arabic{equation}}   
\setcounter{equation}{0}
The concept of symmetry, and its associated algebraic structures, 
is central in the understanding  of the properties of physical systems.
This means, in particular,  that a better  comprehension of the laws of 
physics may  be 
achieved by an identification  of the possible mathematical structures
as well as their classification. For instance, the
properties of elementary particles and their interactions  are very well 
understood  within  Lie algebras. Moreover, the
discovery of supersymmetry gave rise to the concept of Lie
superalgebras which becomes central in theoretical physics and mathematics.
Of course not all the mathematical structures would be  relevant in physics.
For instance,  they  are constraint by the 
principle of quantum mechanics and relativity. This was synthesize in
two no-go theorems which restrict drastically the possible Lie algebras
\cite{cm} and Lie superalgebras \cite{hls} one is able to consider in 
physics. 

But it turns out that Lie (super)algebras are not the only allowed
structures one is able to consider. Several attempts to construct 
models based on different algebras were  proposed.
Here we focus on one of the possible extensions called fractional
supersymmetry \cite{ FSUSY4,FSUSY6,  FSUSY2, FSUSY5,FSUSY,FSUSY3, cubic1,cubic3,  cubic2,FSUSY7, 
FSUSY8, FSUSY9, cubic4} together with its associated 
underlying algebraic structure named $F-$Lie algebra or
Lie algebra of order $F$  \cite{Michel-finit, Michel-infinit} (note that a different approach 
has been proposed in \cite{Kerner}).
Lie algebras of order $F$ lead to  new models of symmetry of space-time 
{\it i.e.} lead to some new non-trivial extensions of the Poincar\'e
algebra, which involves
$F-$arry ($F \ge 3$) relations instead of the
 usual quadratic ones \cite{cubic1, cubic3,cubic2,cubic4}.
These new structures can be seen as a
possible  generalization of Lie (super)algebras.
An $F-$Lie algebra admits a $\mathbb Z_F-$gradation, the zero-graded part
being a Lie algebra. An $F-$fold symmetric product (playing the role of
the anticommutator in the case $F=2$)  expresses the zero graded  part
in terms of the non-zero graded part.

Subsequently, a detailed analysis when  $F=3$ and for a specific extension
of the Poincar\'e algebra was  undertaken together with its explicit
implementation in 
quantum field theory \cite{cubic1, cubic3, cubic2,cubic4}. However, this general study reveled
some difficulties that are not already resolved. This means that in order
to understand the impact of these new structures, a general
algebraic  study  should be undertaken. Thus, the aim  of this paper is 
two-fold.
Firstly a general classification of Lie algebra of order $3$ is
initiated when the zero graded part of the algebra is either
(i) $\s$ or (ii) the Poincar\'e algebra in  four dimensions
$\mathfrak{iso}(1,3)$.  It is then shown that the structure of
the algebra is relatively rigid and a few examples of Lie
algebras of order $3$  are possible in the former case 
(see Theorem \ref{sl2-irrep}).
Although in the latter cases,
since the generators of the space-time translation commute,
there are many possible extensions of
the Poincar\'e algebra (see Theorem \ref{Fpoincare})
Secondly, a  theory of deformations and  contractions is presented. 
This can be seen as a natural extension of
the theory of contraction/deformation of Lie (super)algebras 
(see for example \cite{magnificul, magnificul3}) to 
Lie algebras of order 3. Indeed, contraction/deformation are
relevant in physics in the sense that they may provide a relationship
between two different theories. For instance,
the Poincar\'e algebra  of special relativity and the Galiean
algebra of non-relativistic physics are related through
an In\"on\"u-Wigner contraction. In the same way it is known that
the $N-$extended supersymmetric extension of the Poincar\'e algebra can be
obtained through a contraction of the superalgebra $\mathfrak{osp}(4|N)$.
Similarly the extension of the Poincar\'e algebra studied in
\cite{cubic1, cubic3,cubic2,cubic4} was obtained through a 
contraction of a certain
Lie algebra of order 3 \cite{Michel-finit}.

The content of the paper is organised as follow. In the next section the
definition of Lie algebras of order $F$  is recalled.  Explicit examples
are then given. 
In section three a classification of all Lie algebras of order 3
associated to $\s$ is given. 
 Section four  is devoted to a
general study of Lie algebras of order 3 associated to the
 Poincar\'e algebra in four dimensions $\mathfrak{iso}(1,3)$.
In  section five  the  general notion
(in a topological sense) of contractions  is defined.
This notion  being too general more useful contractions (In\"on\"u-Wigner contractions) are introduced. 
Section six  is devoted to  the implementation  of
the theory of deformations  of Lie algebra of order 3 in the
Gerstenhaber sense. Infinitesimal and isomorphic deformations are  then
introduced.

\section{Lie algebras of order $3$}
\renewcommand{\theequation}{2.\arabic{equation}}   
\setcounter{equation}{0}

In this section we recall the definition and some basic properties of Lie algebras of order 
$F$ introduced in \cite{Michel-finit} and \cite{Michel-infinit}
 and we define the algebraic variety of these algebraic structures.

\subsection{Definition and examples of elementary Lie algebras of order $3$}
\label{3-def}

\begin{definition}
\label{elementary}
Let $F\in\mathbb{N}^*$.
A ${\mathbb Z}_F$-graded ${\mathbb C}-$vector space  ${\mathfrak{g}}= {\mathfrak{g}}_0 \oplus
{\mathfrak{g}}_1\oplus
{\mathfrak{g}}_2 \dots \oplus
{\mathfrak{g}}_{F-1}$ 
is called a complex Lie algebra of order $F$ if
\begin{enumerate}
\item $\mathfrak{g}_0$ is a complex Lie algebra.
\item For all $i= 1, \dots, F-1 $, $\mathfrak{g}_i$ is a representation
of $\mathfrak{g}_0$.
If $X \in \g_0, Y \in \g_i$ then $[X,Y]$ denotes
the action of $X \in \g_0$ on
$Y \in \g_i$ for all $i=1,\cdots F-1$.
\item  For all $i=1,\dots,F-1$ there exists  an $F-$linear,  $\mathfrak{g}_0-$equivariant  map
 $$\mu_i : {\cal S}^F\left(\mathfrak{g}_i\right)
\rightarrow \mathfrak{g}_0,$$
 where  ${ \cal S}^F(\mathfrak{g}_i)$ denotes
the $F-$fold symmetric product of $\mathfrak{g}_i$, satisfying the following (Jacobi) identity
\beqa
\label{eq:J}
&&\sum\limits_{j=1}^{F+1} \left[ Y_j,\mu_i ( Y_1,\dots,
Y_{j-1},
Y_{j+1},\dots,Y_{F+1}) \right] =0, \nonumber
\eeqa
\noi
for all $Y_j \in \mathfrak{g}_i$, $j=1,..,F+1$. 
\end{enumerate}
\end{definition}

\begin{remark}
If $F=1$, by definition $\mathfrak{g}=\mathfrak{g}_0$ and a Lie algebra of order $1$ is a Lie algebra. 
If $F=2$, then $\g$ is a Lie superalgebra. Therefore,  Lie algebras of order $F$ appear as some 
kind of generalisations of Lie algebras and superalgebras.
\end{remark}

Note that by definition 
the following Jacobi identities are satisfied by a Lie algebra of order $F$:
\begin{enumerate}
\item[] For any $ X,X',X'' \in \mathfrak{g}_0$, 

\beq
\left[\left[X,X'\right],X''\right] +  
\left[\left[X',X''\right],X\right] +
\left[\left[X'',X\right],X'\right] =0.
\tag*{J1} \ 
\eeq

\item[] For any $X,X' \in {\mathfrak{g}}_0 \mbox{ and } Y \in {\mathfrak{g}}_i, i=1,..,F-1$,
\beq
 \left[\left[X,X'\right],Y\right] +
\left[\left[X',Y\right],X\right] +
\left[\left[Y,X\right],X'\right]=0.
\tag*{J2} \ 
\eeq
\item[] For any $X\in {\mathfrak{g}}_0 \mbox{ and } Y_j \in {\mathfrak{g}}_i$, 
$j=1,\dots,F$, $i=1,\dots,F-1$,

\beq
\left[X,\mu_i (Y_1,\dots,Y_F)] =
\mu_i (\left[X,Y_1 \right],\dots,Y_F)  +
\dots +
\mu_i (Y_1,\dots,\left[X,Y_F\right] \right).
\tag*{J3}
\eeq

\item[] For any  $Y_j \in \mathfrak{g}_i$, $j=1,...,F+1,\  i=1,...,F-1$,

\beq
\label{exx}
\sum\limits_{j=1}^{F+1} \left[ Y_j,\mu_i ( Y_1,\dots,
Y_{j-1},
Y_{j+1},\dots,Y_{F+1}) \right] =0.
\tag*{J4}
\eeq
\end{enumerate}

\begin{proposition}
Let ${\mathfrak{g}}={\mathfrak{g}}_0\oplus{\mathfrak{g}}_1\oplus\dots\oplus{\mathfrak{g}}_{F-1}$ 
be a Lie algebra of order $F$, 
with $F>1$. For any $i=1,\ldots,F-1$, the $\ZZ_F-$graded vector spaces ${\mathfrak{g}}_0\oplus{\mathfrak{g}}_i$ 
is  a Lie algebra of order $F$. We  call these type of algebras \textit{elementary Lie algebras of order $F$}.
\end{proposition}

\medskip

In \cite{Michel-finit} an inductive process for the construction of Lie algebras of order $F$ starting 
from a Lie algebra of order $F_1$ with
$1\le F_1<F$ is given. In this paper we are especially concerned by deformation and classification problems. 
Moreover, 
we restrict ourselves to elementary Lie algebras of order $3$,
 ${\mathfrak{g}}={\mathfrak{g}}_0\oplus{\mathfrak{g}}_1$.
We also denote the $3-$linear map $\mu_1$ by the $3-$entries bracket $\{.,.,.\}$ and we refer to it as 
a $3-$bracket.
Non-trivial examples of Lie algebras of order $F$ (finite and 
infinite-dimensional) are given in 
\cite{Michel-finit} and  \cite{Michel-infinit}. We now give some examples of finite-dimensional 
Lie algebras of order $3$, which will be relevant in the sequel.

\begin{example}
\label{so23}
Let $\g_0={\mathfrak{so}}(2,3)$ and $\g_1$ its adjoint representation.  Let $\{J_a, a=1,\cdots,10\}$
 be a basis of $\g_0$ 
and $\{A_a={\rm ad}(J_a),a=1,\cdots,10\}$ be the
 corresponding basis of $\g_1$. Thus, one has $[J_a, A_b]={\rm ad}([J_a, J_b])$. 
Let $g_{ab}=Tr(A_aA_b)$ be the Killing form. 
Then one can endow $\g_0\oplus\g_1$ with a 
Lie algebra of order $3$ structure given by
\beq
\{ A_a,A_b,A_c \}=g_{ab}J_c+g_{ac}J_b+g_{bc}J_a. \nonumber
\eeq
\end{example}

\begin{example}
\label{FP}
Let $\g_0$ be the Poincar\'e algebra in four dimensions and  
$\{ L_{mn}, P_m:\ L_{mn}=-L_{nm},\ m<n, m,n,=0,\dots,3 \} $  be a basis of 
$\g_0$ 
with the non-zero brackets
\beqa
 \left[L_{mn}, L_{pq}\right]&=&
\eta_{nq} L_{pm}-\eta_{mq} L_{pn} + \eta_{np}L_{mq}-\eta_{mp} L_{nq},\nonumber \\
\left[L_{mn}, P_p \right]&= &\eta_{np} P_m -\eta_{mp} P_n.
\eeqa
\noi
Let now $\g_1$ be the $4-$dimensional vector representation of $\g_0$, the action
of $\g_0$ on $\g_1$ is given by
$$\left[L_{mn}, V_p \right]= \eta_{np} V_m -\eta_{mp} V_n, \ \
\left[P_{m}, V_n \right]= 0 $$
where  $\{ V_m:\ m=0,\dots,3\}$ is a basis of $\g_1$.
The following brackets on $\g_1$ 
\beqa
\label{algebra0}
&&\{ V_m, V_n, V_r \}=
\eta_{m n} P_r +  \eta_{m r} P_n + \eta_{r n} P_m, 
\eeqa
\noi
with the metric $\eta_{mn}=\rm{diag}(1,-1,-1,-1)$ 
endow $\g_0 \oplus \g_1$ with 
an elementary Lie algebra of order $3$ structure.
\end{example}

\subsection{The variety elementary ${\cal F}_{m,n}$ of Lie algebras of order $3$}

Let $\g=\g_0\oplus\g_1$ be an elementary Lie algebra of order $3$ and let 
$A=(\g_0\otimes \g_0) \oplus (\g_0 \otimes \g_1) \oplus {\cal S}^3 (\g_1)$.
The multiplication of Lie algebra of order $3$  is given by the linear map
$$\varphi : A \to \g$$
satisfying the conditions { J1-J4}.
Let $\varphi_1, \varphi_2, \varphi_3$ be the restrictions of $\varphi$ to 
each of the terms of $A$
$$\varphi_1: \g_0 \otimes \g_0 \to \g_0, $$
$$\varphi_2: \g_0  \otimes \g_1 \to \g_1, $$
$$\varphi_3:  {\cal S}^3 (\g_1)\to \g_0. $$
We denote this by $\p=(\p_1, \p_2, \p_3)$.

Let $\{X_i: i=1, \dots , m\}$ and resp.
$\{Y_a: a=1, \dots , n\}$ be a basis of $\g_0$ and resp. $\g_1$. 
The maps $\p_i$ ($i=1,2,3$)  are defined  by their structure constants
with regard to this basis
\beqa
\label{3-constante}
\p_1 (X_i, X_j)= C_{ij}^k X_k,\ \
 \p_2 (X_i, Y_b)=D_{ib}^c Y_c\mbox{ and } \p_3 (Y_a, Y_b, Y_c)=E_{abc}^i X_i.
\eeqa
\noi
The structure constants $(C_{ij}^k, D_{ib}^c, E_{abc}^i)$ verify the following 
conditions:
\beqa
\label{cond-const}
&&C_{ij}^k=-C_{ji}^k,\ \nonumber \\ 
&&E_{abc}^i=E_{acb}^i=E_{bac}^i=E_{bca}^i=E_{cab}^i=E_{cba}^i.
\eeqa
\noi
and the polynomial equations corresponding to the Jacobi conditions J1-J4 are:
\beqa
\label{jacobi-constante}
&&C_{ij}^\ell C_{\ell k}^m + C_{jk}^\ell C_{\ell i}^m+C_{ki}^\ell C_{\ell j }^m=0, \nonumber\\
&&C_{ij}^k D_{ka}^c-D_{ja}^bD_{ib}^c + D_{ia}^bD_{jb}^c=0, \\
&&E_{abc}^j C_{ij}^k - D_{ia}^d E_{dbc}^k - D_{ib}^d E_{adc}^k - D_{ic}^d E_{abd}^k=0, \nonumber \\
&&D_{ia}^\ell E_{bcd}^i+ D_{ib}^\ell E_{cda}^i + D_{ic}^\ell E_{dab}^i+ D_{id}^\ell E_{abc}^i =0.\nonumber
\eeqa
\noi
Let ${\CC}^N$ be the vector space whose elements are the $N-$tuple 
$(C_{ij}^k, E_{ij}^k, D_{ijk}^l)$, 
with $N=mC_m^2+mn^2+m C_n^3$.
The polynomial equations  \eqref{jacobi-constante} determine 
an algebraic variety ${\cal F}_{m,n}$ embedded in  ${\CC}^N$. Each point of   ${\cal F}_{m,n}$ 
correspond to an $(m+n)-$dimensional Lie algebra of order $3$. 
Thus we identify any elementary 
Lie algebra of order $3$ with the bracket $\p$ to a point 
of  ${\cal F}_{m,n}$.

\medskip

Let us now consider the action of the group $GL({m,n})\cong GL(m)\times GL(n)$ on  ${\cal F}_{m,n}$.
 For any $(h_0,h_1)\in GL(m,n)$, this action is defined by
$$ (h_0, h_1) \cdot (\p_1 ,\p_2 ,\p_3) \to (\p'_1 , \p'_2 , \p'_3) $$
where 
\beqa
\p'_1 (X_1, X_2)&=&h_0^{-1} \p_1 (h_0 (X_1), h_0 (X_2)), \nonumber \\
\p'_2 (X_1, Y_2)&=&h_1^{-1} \p_2 (h_0 (X_1), h_1 (Y_2)), \nonumber \\
\p'_3 (Y_1, Y_2, Y_3) &=& h_0^{-1} \p_3 ( h_1 (Y_1), h_1 (Y_2), h_1 (Y_3)),
\eeqa
\noi
where $(Y_1,Y_2,Y_3)$ represents an element of $\Sy^3(\g_1)$.
The group $GL(m,n)$ can be embedded in $GL(n+m)$. It corresponds to the subgroup of $GL(m+n)$ which let invariant
the subspaces $\g_0$ and $\g_1$ of $\g$.
Denote by ${\cal O}_\p$ the orbit of $\p=(\p_1,\p_2,\p_3)$ with respect to this action. 
Then the algebraic variety ${\cal F}_{m,n}$ is fibered by theses orbits. The quotient set 
is the set of isomorphism  classes of $(m+n)-$dimensional elementary Lie algebras of order $3$.

\section{ $\s$-algebras of order $3$}
\label{3-sl2}
\renewcommand{\theequation}{3.\arabic{equation}}   
\setcounter{equation}{0}

In this section we study complex Lie algebras of order $3$, $\g_0\oplus \g_1$ for which $\g_0\cong \s$
and $\g_1$ is an arbitrary representation of $\s$. 
We denote by $X_+, X_-, X_0$ a standard basis of $\g_0$
\beqa
\label{3-g0}
[X_0, X_+]=2X_+,\ [X_0,X_-]=-2X_-,\ [X_+, X_-]=X_0, 
\eeqa
and ${\cal D}_\ell$ ($\ell \in \mathbb N$) an irreducible representation of dimension $\ell+1$.

\begin{theorem}
\label{sl2-irrep}
The graded complex vector space
$\g\cong\s\oplus \g_1$, with $\g_1$ a  representation of $\s$ is provided with a non-trivial  
Lie algebra structure of order $3$  if and only if:

\begin{itemize}
\item[1.] $\g_1 \cong \D_2$ ($\D_2=\langle Y_2,Y_0,Y_{-2}\rangle$), with the  non-zero three-brackets 
\beqa 
\begin{array}{llllll}
\label{sl2+adj}
\{ Y_{2}, Y_{-2}, Y_0\}&=& X_0,\ &\{ Y_0, Y_0, Y_0\}&=& 6X_0, \\
\{Y_{2}, Y_{-2}, Y_{2}\}&=& 2 X_+,\ &\{Y_{2}, Y_0, Y_0 \}&=& 2 X_+,\\
\{Y_{-2}, Y_{2}, Y_{-2}\}&=& 2 X_-,\ &\{Y_{-2}, Y_0, Y_0 \}&=& 2 X_-.
\end{array}
\eeqa
\item[2.] $\g_1 \cong D_2 \oplus D_0^{(1)} \oplus \cdots \oplus \D_0^{(k)}$
($\D_2=\langle Y_2,Y_0,Y_{-2}\rangle, \D_0^{(k)}=\langle \lambda_k \rangle$), with the non-zero three-brackets 

\beqa
\label{sl2+red}
\{ \lambda_i, \lambda_j, Y_2\} &=& \alpha_{ij} X_+,\nonumber\\
\{ \lambda_i, \lambda_j, Y_0\} &=& \alpha_{ij} X_0, \\
\{ \lambda_i, \lambda_j, Y_{-2}\} &=& \alpha_{ij} X_-.\nonumber
\eeqa
and $\alpha_{ij} \in \CC$.
\item[3.]    $\g_1 \cong \D_1\oplus \D_0$ ($\D_1=\langle Y_1,Y_{-1}\rangle, \D_0=\langle \lambda \rangle $), with the  non-zero three-brackets  
\beqa
\label{sl2+spin}
\{ \lambda, Y_1, Y_1\} &=& -2 X_+,\nonumber\\
\{ \lambda, Y_1, Y_{-1}\} &=&  X_0, \\
\{ \lambda, Y_{-1}, Y_{-1}\} &=& 2  X_-.\nonumber
\eeqa
\end{itemize}
\end{theorem}
{\it Proof.} 
Since $\g_1=\bigoplus \limits_k {\cal D}_{\ell_k}$, with $ {\cal D}_{\ell_k}$ an irreducible representation of dimension 
$\ell_k +1 $, the $3-$brackets $\{ \g_1, \g_1, \g_1 \}$ contain terms like (i) 
$\{ {\cal D}_{\ell_1},  {\cal D}_{\ell_1},  {\cal D}_{\ell_1} \}$, 
(ii) $\{ {\cal D}_{\ell_1},  {\cal D}_{\ell_1},  {\cal D}_{\ell_2} \}$,
(iii) $\{ {\cal D}_{\ell_1},  {\cal D}_{\ell_2},  {\cal D}_{\ell_3} \}$.

\medskip

\qquad I. Consider firstly the case $\{ {\cal D}_{\ell},  {\cal D}_{\ell},  {\cal D}_{\ell} \}$.
A simple weight argument shows that $\ell$ is even, furthermore the non-vanishing  three brackets are

$$\left\{Y_i,Y_j,Y_k\right\}=
\left\{
\begin{array}{ll}
\alpha_{ijk}X_+&i+j+k=2, \\
\beta_{ijk}X_0&i+j+k=0,\\
\gamma_{ijk}X_-&i+j+k=-2.
\end{array} \right.
$$
Suppose firstly that $\ell= 2$.
The action of $\s$ on $\D_2$ is
\beqa
\begin{array}{lll}
\left[X_0,Y_{-1}\right]=-2Y_{-1},&\left[X_-,Y_0\right]=2Y_{-1},&\left[X_-, Y_{1}\right]=-Y_0,\\
\left[X_+, Y_{-1}\right]=Y_0,&\left[X_+,Y_0\right]=-2Y_{1},&\left[X_0, Y_{1}\right]=2Y_{1}.
\end{array}
\eeqa
\noi
>From symmetry considerations one has  $\alpha_{ijk}=\gamma_{-i,-j,-k}$, and the Jacobi
identity J3 gives
$\alpha_{1,1,-1}=  \alpha_{1,0,0}=\gamma_{-1,-1,1}=  \gamma_{-1,0,0}=2t, \beta_{1,-1,0}=t, \beta_{0,0,0}= 6 t,$ with $t \in \mathbb C$.
Furthermore, a direct calculus shows that the Jacobi identities J4 are satisfied for any $t$.
If $t=0$, the Lie algebra of order $3$ is trivial. If $t \ne 0$ all the algebras are equivalent.
Since  for $\s$ the Casimir operator is given by $Q=\frac12H^2+X_+X_-+X_-X_+$  we have 
$\text{Tr}(X_+ X_-)=g_{+-}=g_{-+}=1, \text{Tr}(X_0 X_0)=g_{00}=2$ and the three-brackets 
\eqref{sl2+adj} can be rewritten \cite{Michel-finit}

$$\left\{Y_i,Y_i,Y_k\right\}=g_{ij}X_k +g_{jk}X_i +g_{ki}X_j,$$
(here $X_+,X_0,X_-$ are denoted $X_{2},X_0,X_{-2}$).
\medskip

\noindent Now we assume that $\ell > 2$ and we prove that
$\{\D_\ell,\D_\ell,\D_\ell \} =0$.
\begin{enumerate}
\item
The bracket $\{Y_0,Y_0,Y_0\}=0$ as  a consequence of J4 applied to $(Y_\ell,Y_0,Y_0,Y_0)$.

\item
The bracket $\{Y_0,Y_0,Y_{i}\}=0$ as a consequence of J4 applied to $(Y_0,Y_0,Y_0,Y_i)$.

\item
The bracket $\{Y_0,Y_i,Y_{j}\}=0,\ i+j \ne 0$ as a consequence of  J4 applied to $(Y_0,Y_0,Y_i,Y_j)$.

\item
The bracket $\{Y_0,Y_i,Y_{-i}\}=0$. If $i\ne 2$ this is a consequence of  J4 applied to $(Y_0,Y_i,Y_i,Y_{-i})$
and if $i=2$  this is a consequence of  J4 applied to $(Y_\ell,Y_0,Y_2,Y_{-2})$.

\item
The bracket $\{Y_i,Y_j,Y_{k}\}=0,\ i+j+k\ne 0$ as a  consequence of J4 applied to $(Y_0,Y_i,Y_j,Y_k)$.

\item
The bracket $\{Y_i,Y_j,Y_{-i-j}\}=0$  as a  consequence of J4 applied to $(Y_k,Y_i,Y_j,Y_{-i-j})$ with $k\ne i,j, -i-j,0$. Such a $k$ 
always exists since the dimension of ${\cal D}_\ell$ is bigger than $5$.
\end{enumerate}

\qquad II. Consider now the case  $\{ {\cal D}_{\ell_1},  {\cal D}_{\ell_1},  {\cal D}_{\ell_2} \}$.

\noi
$\bullet $ We prove now that $\{ {\cal D}_{\ell},  {\cal D}_{\ell},  {\cal D}_{0} \}=0$ for all $\ell$ but $\ell=1$.

\begin{enumerate}
\item
It is easy to see that  $\ell = 0$ leads to a trivial Lie algebra of order $3$.
\item
If $\ell=1$ we denote $ {\cal D}_0=<\lambda >$ and $ {\cal D}_1=<Y_1, Y_{-1} >$. The action of $\s$ on
$\D_1$ is

$$
\begin{array}{ll}
\left[X_0,Y_1\right]=Y_1,&\left[X_+,Y_{-1}\right]=Y_1, \\
\left[X_-,Y_1\right]=Y_{-1},&\left[X_0,Y_{-1}\right]=-Y_{-1},
\end{array}
$$
and a simple weight argument gives

$$
\lbrace \lambda, Y_1,Y_1 \rbrace =\alpha X_+, \ \
\lbrace \lambda, Y_1,Y_{-1} \rbrace =\beta X_0, \ \ 
\lbrace \lambda, Y_{-1},Y_{-1} \rbrace =\gamma X_-.
$$
The Jacobi identities J3 and J4 give $\alpha= -2 t, \beta=t, \gamma = 2t, \ t \in \mathbb C.$
\item
If $\ell =2 $ using the Clebsch-Gordan decomposition ${\cal D}_2 \otimes {\cal D}_2 = {\cal D}_4 \oplus {\cal D}'_2 \oplus 
{\cal D}_0$, 
since the representation ${\cal D}'_2$ is antisymmetric in the permutation of the two factors ${\cal D}_2$, 
we have $\{ {\cal D}_2, {\cal D}_2, {\cal D}_0 \}=0$.
\item
If $\ell >2$, we denote $ {\cal D}_\ell=<Y_\ell, Y_{\ell -2}, \dots, Y_{-\ell} >$.
A simple weight arguments shows that the possible non-vanishing $3-$brackets are: 
$\{ \lambda, Y_i, Y_{-i+2} \},\, \{ \lambda, Y_i, Y_{-i} \}$ and $\{ \lambda, Y_i, Y_{-i-2} \}$.

\begin{enumerate}
\item
The brackets $\{ \lambda, Y_i, Y_{-i+2} \}=0 $, $i\ne 0,-2, \ell$  as  a consequence of J4 applied to $( \lambda, Y_i, Y_{i}, Y_{-i+2})$.

\item
The bracket $\{ \lambda, Y_0, Y_{2} \}=0 $  as  a consequence of J4 applied to $(\lambda, Y_0, Y_{2}, Y_{2}). $

\item 
The  brackets $\{ \lambda, Y_{-2}, Y_{4} \}=0 $  as  a consequence of J4 applied to 
$( \lambda, Y_{-2}, Y_{-2}, Y_{4}). $

\item
The brackets $\{ \lambda, Y_\ell, Y_{-\ell+2} \}=0 $  as  a consequence of J4 applied to 
$( \lambda, Y_\ell, Y_{-\ell+2}, Y_{-\ell+2}). $

\item
The brackets $\{ \lambda, Y_i, Y_{-i} \}=0 $, $i\ne 0$  as  a consequence of J4 applied to $(\lambda, Y_i, Y_{i}, Y_{-i})$.

\item
The bracket $\{ \lambda, Y_0, Y_{0} \}=0 $,   as  a consequence of J4 applied to $(\lambda, Y_0, Y_0, Y_{\ell})$.

\item
The brackets $\{ \lambda, Y_i, Y_{-i-2} \}$ go along the same line as the brackets $\{\lambda, Y_i, Y_{-i+2})$.
\end{enumerate}
\end{enumerate}

\medskip

\noi
$\bullet $
We now consider the brackets of type $\{ {\cal D}_{\ell},  {\cal D}_{0},  {\cal D}_{0} \}$.
By a weight argument and identity J4 one has  $\ell = 2$ and the non-trivial
$3-$brackets are:
\beqa
\{ \lambda, \lambda, Y_2\} &=& \alpha X_+,\nonumber\\
\{ \lambda, \lambda, Y_0\} &=& \alpha X_0,\nonumber\\
\{ \lambda, \lambda, Y_{-2}\} &=& \alpha X_-.\nonumber
\eeqa
\noi    

\noi
$\bullet $ 
We now prove that $\{ \D_{\ell_1}, \D_{\ell_1}, \D_{\ell_2}\}=0$ 
with $\ell_1,\ell_2 \ne 0$.

\begin{enumerate}
\item Let  $\ell_1 =\ell_2 =2$, we  denote 
$\D_2 =\langle Y_2,Y_0,Y_{-2}\rangle, 
\D'_2 =\langle Y'_2,Y'_0,Y'_{-2}\rangle$ the two three-dimensional 
representations. 
In this case, we have four types of brackets :
$\{\D_2,\D_2,\D_2\}, \ \{\D'_2,\D'_2,\D'_2\}, \{\D_2,\D_2,\D'_2\},  
\{\D_2,\D'_2,\D'_2\}$. The possible non-vanishing three-brackets 
are

\beqa
\{Y_i,Y_j,Y_k\}=  \alpha g_{ij} X_{k} + \alpha g_{jk} X_{i}+ \alpha g_{ki} X_{j},
\ \ \ \ \ \ \ \ \ \ \ \ \ \ \ \ \ \ \ \ \ \ \ \ \ \ \ \ \ \ \ \ \ \ \ \ \ \ \ \ \ \ \ \ \   \nonumber  \\
\{Y'_i, Y'_j,Y'_k\}=\alpha' g_{ij} X_{k} + \alpha' g_{jk} X_{i}+ \alpha' g_{ki} X_{j}
\ \ \ \ \ \ \ \ \ \ \ \ \ \ \ \ \ \ \ \ \ \ \ \ \ \ \ \ \ \ \ \ \ \ \ \ \ \ \ \ \ \ \ 
\nonumber \\
\begin{array}{llll}
\{Y_2,Y_2,Y'_{-2}\}=\alpha_1 X_+&\{Y'_{2},Y_{2},Y_{-2}\}=\alpha_2 X_{+}&
\{Y_2,Y_0,Y'_0\}=\alpha_3 X_+&\{Y'_2,Y_0,Y_0\}=\alpha_4 X_+ \\
\{Y_{-2},Y_{-2},Y'_{2}\}=\alpha_1 X_-& \{Y'_{-2},Y_{-2},Y_{2}\}=\alpha_2 X_{-}&
\{Y_{-2},Y_0,Y'_0\}=\alpha_3 X_-&\{Y'_{-2},Y_0,Y_0\}=\alpha_4 X_+ \\
\{Y_2,Y_{-2}Y'_0\}=\beta_1 X_0&\{Y_2,Y_0,Y'_{-2}\}=\beta_2 X_0&
\{Y'_2,Y_{-2},Y_0\}=\beta_3 X_0&\{Y_0,Y_0,Y'_0\}=\beta_4 X_0
\end{array}
\eeqa

\noi
(plus similar terms with $\{Y',Y',Y\}$ and coefficients $\alpha_1',\cdots, \beta_4'$,
as $ \{Y'_2,Y'_2,Y_{-2}\}=\alpha'_1  X_+$, {\it etc.}).
Working out the Jacobi identity J4, one gets that the coefficients $\alpha, \alpha_1,\cdots,
\beta_4,\alpha',\alpha_1',\cdots, \beta_4'$ must be zero. In fact one has
\begin{enumerate}
\item  $(Y_2,Y_2,Y'_{-2},Y'_{-2})$ gives $\alpha_1=\alpha'_1=0$;
\item  $(Y_{-2},Y_{-2},Y_2,Y'_2)$ gives $\alpha=\alpha_2=0$ (resp.  $\alpha'=\alpha_2'=0$);
\item  $(Y'_0,Y'_0,Y_0,Y_2)$ gives $\alpha_3=0$ (resp.  $\alpha'_3=0$);
\item  $(Y_0,Y_0,Y_0,Y'_2)$ gives $\alpha_4=0$ (resp.$\alpha'_4=0$);
\item $(Y_2,Y_2,Y_{-2},Y'_0)$ gives $\beta_1=0$ (resp.$\beta_1'=0$);
\item $(Y_0,Y_0,Y_{2},Y'_{-2})$ gives $\beta_2=0$ (resp.$\beta_2'=0$);
\item $(Y_{-2},Y_{0},Y'_2,Y'_2)$ gives $\beta_3=0$(resp. $\beta_3'=0$); 
\item $(Y_{2},Y_{0},Y_0,Y'_0)$ gives $\beta_4=0$ (resp. $\beta_4'=0$).
\end{enumerate}

Thus all the brackets vanishes.

\item $\ell_1\ne 2, \ell_2$ arbitrary 
${\cal D}_{\ell_1} = < Y_{\ell_1}, Y_{\ell_1 - 2}, \dots, Y_{-\ell_1}>$ and 
$\D_{\ell_2} = < Y'_{\ell_2}, Y'_{\ell_2 - 2}, \dots, Y'_{-\ell_2}>$. 
We consider the bracket 
$\{ Y_i, Y_j, Y'_k \}$.
The identity J4  applied to $( Y_i,  Y_i, Y_j, Y'_k)$ leads to the 
vanishing of the $3-$brackets 
except when  $2i+k\ne 0, \pm 2$ or ($i\ne \ell$ and $ i+j+k\ne 2$).
The identity J4  applied to $( Y_i,  Y_j, Y_j, Y'_k)$ leads to the 
vanishing of the $3-$brackets 
except when  $2j+k\ne 0, \pm 2$ or ($j\ne \ell$ and $ i+j+k\ne 2$). The cases 
that remains to be studied
are:

\begin{enumerate}
\item
If  $k=-2j +2 = -2i +2$ then $i=j$ and the bracket $\{Y_i, Y_i, Y'_{-2i+2}\}$ vanishes.

\item
If  $k=-2j +2 = -2i$ then $i=j-1$ which is not possible.

\item
If $k=-2j+2=-2i-2$ then $j=i+2$ and the bracket reduces to $\{Y_i, Y_{i+2}, Y'_{-2i-2}\}$. 
Then identity J4  applied to $( Y_{\ell_1}, Y_i, Y_{i+2}, Y'_{-2i-2})$ and to $( Y_{-\ell_1}, Y_i, Y_{i+2}, Y'_{-2i-2})$
leads to $\{Y_i, Y_{i+2}, Y'_{-2i-2}\}=0.$

\item
If  $k=-2j=-2i$ then $i=j$ and the bracket vanishes  as  before. If $k=-2j=-2i-2$ then $j=i+1$ which is also excluded.
If $k=-2j-2=-2i-2$ then $i=j$  and the bracket vanishes  as  before.

\end{enumerate}

\item Let $\ell_1=2, \ell_2 \ne 2$ and consider the brackets of the type $\{D_2,\D_2,\D_{\ell_2}\}$.
The Jacobi identity J4 applied on $(Y_{i_1},Y_{i_2},Y'_{i_3},Y'_{i_3})$ with $Y_{i_1},Y_{i_2} \in \D_2$ and $Y_{i_3} \in \D_3$ leads to
$\{D_2,\D_2,\D_{\ell_2}\}=0$.
\end{enumerate}

\qquad III. Consider now the case $\{ \D_{\ell_1}, \D_{\ell_2}, \D_{\ell_3}\}$.

\noi If $\ell_1=\ell_2=0$, by weight arguments we have $\ell_3=2$ and the possible non-vanishing three brackets are

\beqa
\{ \lambda_1, \lambda_2, Y_2\} &=& \alpha_{12} X_+,\nonumber\\
\{ \lambda_1, \lambda_2, Y_0\} &=& \alpha_{12} X_0,\nonumber\\
\{ \lambda_1, \lambda_2, Y_{-2}\} &=& \alpha_{12} X_-.\nonumber
\eeqa

\noi    
where $\D_0^{(1)}=<\lambda_1>, \D_0^{(2)}=<\lambda_2>$ and $\D_2=<Y_2,Y_0,Y_{-2}>.$

\noi
$\bullet$ If $\ell_1=0$ and $\ell_2, \ell_3 \ne0$ the Jacobi identity J4 
applied to $(\lambda,Y,Y,Y')$ with  $\lambda \in \D_0, 
Y\in \D_{\ell_2}, Y' \in \D_{\ell_3}$ leads
to $\{\lambda, Y,Y'\}=0$ except if $\ell_2=\ell_3=1$.

\noi
$\bullet$ If $\ell_1=0$ and $\ell_2=\ell_3=1$, denoting
$\D_1=<Y_1,Y_{-1}>, \D_1'=<Y_1',Y'_{-1}>, \D_0=<\lambda >$, 
the non-vanishing brackets are

\beqa
&\{Y_1,Y_1,\lambda\}= -2 \alpha X_+, \ 
\{Y_1,Y_{-1},\lambda\}=  \alpha X_0,
\{Y_{-1},Y_{-1},\lambda\}= 2 \alpha X_-  \nonumber \\
&\{Y'_1,Y'_1,\lambda\}= -2 \alpha' X_+, \ 
\{Y'_1,Y'_{-1},\lambda\}=  \alpha' X_0,
\{Y'_{-1},Y'_{-1},\lambda\}= 2 \alpha' X_- \nonumber   \\
&\{Y_1,Y'_1,\lambda\}= \beta_1 X_+, \ 
\{Y_1,Y'_{-1},\lambda\}=  \beta_2 X_0,\
\{Y'_1,Y_{-1},\lambda\}=  \beta_3 X_0,\
\{Y_{-1},Y'_{-1},\lambda\}= \beta_4 X_-. \nonumber  
\eeqa

\noi The Jacobi identity J4 with $(\lambda, Y_1,Y_1,Y'_{-1})$ implies 
$\alpha=\beta_2=0$, with $(\lambda,Y'_1,Y'_1,Y_{-1})$ implies
$\alpha'=\beta'_2=0$, with $(\lambda,Y_1,Y_{-1},Y'_{1})$ implies
$\beta_1=0$ and with $(\lambda,Y_{-1},Y_{1},Y'_{-1})$ implies $\beta_4=0$.

\noi
$\bullet$ If $\ell_1, \ell_2, \ell_3\ne 0$,
let $Y\in \D_{\ell_1}, Y'\in \D_{\ell_2}, Y''\in \D_{\ell_3}$. Then J4  applied to $( Y,  Y, Y', Y'')$
 leads to $ \{Y, Y', Y''\}=0$. 

Taking all the cases obtained above, the only Lie algebras of order 3 
associated to $\s$ are 
$\g= \s \oplus \D_2$, $\g=\s \oplus D_2 \oplus D_0^{(1)} \oplus \cdots 
\oplus \D_0^{(k)}$
or $\g=\s \oplus \D_0 \oplus \D_1$  with brackets given in
\eqref{sl2+adj}, \eqref{sl2+red} and \eqref{sl2+spin}. QED.

\medskip
\begin{remark}
In \cite{Michel-finit} two families of algebras associated to $\frak{sp}(n)$ 
were constructed.
We have however check that they coincide when $n=1$ {\it i.e.} for $\s$.
This algebra is the one  given in Eq.[\ref{sl2+adj}].
The algebra \eqref{sl2+red} was also obtained in \cite{Michel-finit}  and the algebra
\eqref{sl2+spin} in \cite{Michel-finit}
and \cite{ayu}.
\end{remark}

\section{Poincar\'e algebras of order $3$}
\renewcommand{\theequation}{4.\arabic{equation}}   
\setcounter{equation}{0}
In this section we study and 
provide a systematic method to obtain
all elementaries
Lie algebras of order $3$,  $\g=\g_0 \oplus \g_1$   where
$\g_0$ is isomorphic to the Poincar\'e algebra and $\g_1$ is
an arbitrary finite dimensional representation of the Poincar\'e
algebra. Such  algebras are 
 called {\it Poincar\'e algebras of order $3$}.

\noindent 
To construct these algebras, we proceed in several steps.
Firstly, we extend the action of 
$\mathfrak{so}(1,3,\mathbb C)= \mathfrak{so}(1,3)\otimes_\mathbb R \mathbb C$
 on $\g_1$, with $\g_1$ a finite dimensional
representation of $\mathfrak{so}(1,3,\mathbb C)$, to the action of the 
complexified  Poincar\'e algebra  on
$\g_1$.  Then, we construct the $\mathfrak{so}(1,3,\mathbb C)-$equivariant
 mappings
from $\Sy^3(\g_1)$ into $\D_{1,1}$, with $\D_{1,1}$ the vector
representation of $\mathfrak{so}(1,3,\mathbb C)$. 
Finally, we obtain all Lie algebras
of order $3$, $\g= \big(\mathfrak{so}(1,3,\mathbb C) \oplus
 \D_{1,1}\big) \oplus \g_1$.

\subsection{Finite dimensional representations of the Poincar\'e algebra}

The Poincar\'e algebra  in  $(1+3)$dimensions $\mathfrak {iso}(1,3)$ 
(see Example \ref{FP} for notations) is given by

\beqa
\label{Poincare}
&&\left[L_{mn}, L_{pq}\right]=
\eta_{nq} L_{pm}-\eta_{mq} L_{pn} + \eta_{np}L_{mq}-\eta_{mp} L_{nq}, \nonumber \\
&& \left[L_{mn}, P_p \right]= \eta_{np} P_m -\eta_{mp} P_n,
\ \ \left[P_m, P_n \right]= 0, \ \
\eeqa
\noindent
where $\eta_{mn}$ is the Minkowski metric. Let 
$\mathfrak {iso}(1,3,\CC)= \mathfrak {iso}(1,3) \otimes_\mathbb{R} 
\mathbb{C}$ be the complexified of  $ \mathfrak {iso}(1,3)$.  
Its Levi part is  isomorphic to $\s \oplus \s$.
Consider in \eqref{Poincare} the following change of basis

\beqa
\label{vect}
&\begin{array}{ll}
U_0=i L_{12}-L_{03};&V_0=iL_{12}+L_{03};\\
U_+=\frac12 \left(i L_{23}-L_{31}-L_{01}-i L_{02}\right);&
V_+=\frac12 \left(i L_{23}-L_{31}+L_{01}+i L_{02}\right);
\\
U_-=\frac12 \left(i L_{23}+L_{31}-L_{01}+i L_{02}\right);&
V_-=\frac12 \left(i L_{23}+L_{31}+L_{01}-i L_{02}\right);
\end{array} \nonumber \\
&\begin{pmatrix} p_{+-}& p_{--} \\
                p_{++}&p_{-+}\end{pmatrix}=P_m \sigma^m=
\begin{pmatrix}P_0+P_3&P_1-iP_2 \\
               P_1+iP_3&P_0-P_3\end{pmatrix},
\eeqa

\noi
(with $\sigma^0$ the identity matrix and $\sigma^i, \ i=1,2,3$ the
Pauli matrices).
In this basis the ${\frak{iso}}(1,3, \CC)$ brackets  are given by

\beqa
\label{poincare}
&\begin{array}{ll}
\left[U_0,U_\pm\right]=\pm 2 U_\pm,  & \left[V_0,V_\pm\right]=\pm 2 V_\pm, \\
\left[U_+,U_-\right]=U_0,&\left[V_+,V_-\right]=V_0,
\end{array}
\nonumber \\
&\begin{array}{ll}
\left[U_+, p_{- \varepsilon}\right]=p_{+\varepsilon},
&\left[V_+, p_{ \varepsilon-}\right]=-p_{\varepsilon +}, \\
\left[U_-, p_{+\varepsilon }\right]=p_{-\varepsilon}&
\left[V_-, p_{ \varepsilon +}\right]=-p_{\varepsilon -}, \\
\left[U_0, p_{\varepsilon \varepsilon'}\right]=
\varepsilon p_{\varepsilon \varepsilon'},&
\left[V_0, p_{\varepsilon \varepsilon'}\right]=
\varepsilon' p_{\varepsilon \varepsilon'},
\end{array}
\eeqa

\noi
(with $\varepsilon, \varepsilon'= \pm$).

Let $\D_i$ be the irreducible $(i+1)-$dimensional representation of $\s$.
We note by 
${\cal D}_{i,j} = {\cal D}_i\otimes {\cal D}_j, \ i,j \in \mathbb N$ 
the  irreducible representation (since there 
is a factor 2 in the first line on equation
\eqref{poincare} $i$ belongs to $ \mathbb N$ and not to $\frac12 \mathbb N$)
of 
  dimension $(i+1)(j+1)$ of 
$\s \oplus \s$ defined from

$$ \rho(U_0 + V_0) (x \otimes y)= [U_0,x]\otimes y + x \otimes [V_0,y].$$
Let  $\g_1= \oplus_k {\cal D}_{i_k,j_k} $ be  an arbitrary
reducible representation of $\s \oplus \s.\ $

\begin{lemma}
\label{rep}
Let $\D$ be a  
(finite dimensional) representation of $\s \oplus \s$.
The action of $\s \oplus \s$ on ${\cal D}$ extends to an
action of $\frak{iso}(1,3,\mathbb C)$ on ${\cal D}$ such that:
\begin{enumerate}
\item[1.] The operators  $\rho(P_m)$ 
 ($m=0,\cdots,3$), are nilpotent.
\item[2.] Let

$$A_p= 
\bigcap \limits_{p_0+p_1+p_2+p_3=p} 
\mathrm{Ker }\Big(\left(\rho( P_0)\right)^{p_0}
\left(\rho( P_1)\right)^{p_1}
\left(\rho( P_2)\right)^{p_2}
\left(\rho( P_3)\right)^{p_3}\Big),
 $$
then, there exists an $N$ such that we have the filtration

$$A_1 \subset A_2 \subset \cdots \subset A_N= \D,
$$
and for every $0\le p\le N$,
$A_p$ is an 
$\s \oplus \s$ module.
\item[3.] ${\cal D}$ is indecomposable
({\it i.e.} one can find  an irreducible representation $\D' \subset D$
of $\mathfrak{iso}(1,3)$
such that it is impossible to have $\D = \D' \oplus \D''$ where
$\D''$ is stable under the action $\rho(P_m)$ $-$see Examples \ref{ex-ind}
 below $-$),

$$\xymatrix{
0&A_1
\ar[l]_{\rho( P_m ) }&
A_2 
\ar[l]_{\rho( P_m)  }&
\ar[l]_{  \rho( P_m ) } \ \ \ \cdots  \ \ \ &
A_N
\ar[l]_{\ \ \ \rho( P_m)    }
}=\D.
$$
Let $B_p^m= \rho( P_m)\left(A_p\right)$. Then $B_p^0=B_p^i$ for $i=1,2,3$. We denote by
$B_p$ this space and we have

$$B_p \subset A_p \otimes \D_{1,1} \subset A_{p-1}.$$
\end{enumerate}  
\end{lemma} 

{\it Proof.} \ \  1.
Let $\lambda_0$ be an arbitrary eigenvalue of $\rho(P_0)$ and 
$E_0=\ker\left(\lambda_0-\rho(P_0)\right) \subseteq
\ker\left(\lambda_0-\rho(P_0)\right)^{n_0} \subseteq \D$ 
(with $\ker\left(\lambda_0-\rho(P_0)\right)^{n_0}$ the generalised 
eigenspace). Denote $\lambda_i \ (i=1,2,3)$ an arbitrary  eigenvalue of 
$\rho(P_i){}_{|_{_{ E_0}}}$ and let $V=\ker\left(\lambda_i-\rho(P_i){}_{|_{_{ E_0}}}
\right) 
\subseteq \left(\lambda_i-\rho(P_i){}_{|_{_{ E_0}}}\right)^{n_i}
\subseteq E_0$. 
Since $[\rho(P_i), \rho(P_0)]=0$ then $E_0$ is invariant by the action
of $P_i$.
In $V$ we have  $\rho(P_0){}_{|_{_V}}= \lambda_0 \text{Id}$,
$\rho(P_i){}_{|_{_V}}= \lambda_i \text{Id}$, since $[L_{i0},P_i]=P_0$ we 
have $\lambda_0=\lambda_i=0$. And $\rho(P_0),\rho(P_i)$ are nilpotent operators.

\noi
2. Since the operators $\rho( P_m)$ are nilpotent (we denote $n_m$ the index of
nilpotency of $\rho( P_m)$) and are commuting operators, it is obvious
that there exists an $N \ge \text{max}(n_0,n_1,n_2,n_3)$ such that
 $A_N =\D$. 
Furthermore, since for all $ L \in \s \oplus \s$,   there exists a 
$P \in \D_{1,1}$ such that $[L,P]=P_0$ we have  $[L,P^p]=pP_0 P^{p-1}$. This means that if
$v \in A_p$, 
$[L,v] \in A_p$. Thus,
$A_p$ is an $\s \oplus \s$ module.

\noi
3. Let $w \in B_p^0$,
then there exists $v \in A_p$ such that
$w= [P_0,v]$. Since $A_p$ is an $\s \oplus \s$
module we have $[L_{0i},v']=v$ ($i=1,2,3$) 
with $v' \in A_p$.
Using the Jacobi identity of Lie algebras, $w=[P_0,v]=[P_0,[L_{0i},v']]$ leads to
$[P_i,v']=w -[L_{0i},[P_0,v']]$
and $B_p^i \subseteq B_p^0$.
The converse goes along 
the same lines and we have $B_p^0=B_p^i$ for $i=1,2,3$. Finally, the $\s \oplus \s$ equivariance
of the mapping guarantees that,
$B_p \subset A_p \otimes \D_{1,1} \subset A_{p-1}$
 and
$\D$ is indecomposable. 
QED.

\begin{example}\label{ex-ind}
(1) If $\g_1 = {\cal D}_{1,1} \oplus {\cal D}_{0,0}$ the vector plus the
scalar representations of
$\s \oplus \s$,  $P_m$ can be represented by $5 \times 5$ nilpotent
matrices. 
If we denote $\left<v_m, m=0,\cdots,3\right>$
(resp. $\left<w_0\right>$) a basis of ${\cal D}_{1,1}$ 
(resp.  $ {\cal D}_{0,0}$) we can define

$$ \rho(P_m )w_0= v_m, \ \ \rho(P_m) v_n =0.$$

\noi
(2) If $\g_1 = {\cal D}_{1,0} \oplus {\cal D}_{0,1}$,
 since $\D_{1,0} \otimes \D_{1,1} = \D_{2,1} \oplus \D_{0,1} 
\supset \D_{0,1}$,  
$P_m$ can be represented by the  $4 \times 4$ matrices
$\rho(P_m)=\begin{pmatrix}0&\sigma_m \\0&0\end{pmatrix}$ 
such that for $\psi \in \D_{1,0}, \bar \chi \in \D_{0,1}$  we have

$$ \rho(P_m) \psi=0, \ \ \rho(P_m) \bar \chi = \sigma_m \psi.$$

\noi
(3) The example above can be  even refined. Let
$\g_1=\D_{1,0} \oplus \D_{0,1} \oplus \D_{0,1}'.$ The action of the $P$'s
can be defined as follow:

$$ \rho(P_m) \psi=0,  \rho(P_m) \psi'=0
\ \ \rho(P_m) \bar \chi = \sigma_m \psi,$$
with $\psi \in \D_{1,0},\psi'\in \D'_{1,0},\bar \chi \in \D_{0,1}.$
Here, $\ker(\mathrm{ad~}P_m)= \D_{1,0} \oplus \D_{1,0}',$
 $\ker(\mathrm{ad~}P_m)^2= \D_{1,0} \oplus \D_{0,1} \oplus \D_{0,1}'$ ($m=0,\dots,3)$ and 
$B_2= \D_{1,0}.$

\end{example}

\begin{remark}
\label{irred}
If $\g_1$ is an irreducible representation of $\s \oplus \s$,
then the action of $\ad$ on $\g_1$ is trivial. Indeed,
since $\g_1$ is irreducible, $\ker\left(\mathrm{ad ~}P_m\right)$ is equal
either to $\g_1$ or $\left\{0\right\}$. But since $\ad_0,\cdots,\ad_3$
commute they can be simultaneously diagonalised this means that
$\ker\left(\mathrm{ad ~} P_m\right) \ne \left\{0\right\}$ and the action of $\ad$ on
$\g_1$ is trivial. 
\end{remark} 

\subsection{$\s \oplus \s -$equivariant mappings}
Now, we construct the possible $\s \oplus \s-$equivariant mappings
 from $\Sy^3(\g_1)$ into
$\D_{1,1}$, with $\g_1$ an arbitrary representation of
$\s \oplus \s$.
We recall the following  isomorphisms of representations of
$GL(A) \times GL(B)$ \cite{fulton-harris}:

\beqa
\label{sum}
{\cal S}^p \left(A \oplus B \right)&=&
\bigoplus\limits_{k=0}^p {\cal S}^k \left(A \right)\otimes 
{\cal S}^{p-k} \left(B\right)  \nonumber \\
{\cal S}^p \left(A \otimes B\right)&=& 
\bigoplus \limits_{\Gamma} {\$}^{\Gamma} \left(A\right) \otimes
{\$}^{\Gamma} \left(B\right), 
\eeqa

\noindent
where the second  sum is taken over all Young diagrams $\Gamma$ of length $p$
and ${\$}^{\Gamma}\left(A\right)$ denotes the irreducible representation
of $GL(A)$ corresponding to the Young symmetriser of $\Gamma$.
In particular this gives
\beqa
\label{equiv}
&&{\cal S}^3 \left(A \oplus B \oplus C\right)=
{\cal S}^3 \left( A\right) \ \oplus  \ {\cal S}^3 \left( B \right)
\ \oplus \  {\cal S}^3 \left( C\right) \ \oplus\  A \otimes B \otimes C 
\ \oplus \nonumber \\
&& {\cal S}^2 \left( A\right) \otimes B \ \oplus \
{\cal S}^2 \left( A\right) \otimes C \ \oplus\
{\cal S}^2 \left( B\right) \otimes A \ \oplus \
{\cal S}^2 \left( B\right) \otimes C \ \oplus \
{\cal S}^2 \left( C\right) \otimes A \ \oplus \
{\cal S}^2 \left( C\right) \otimes B   \\
&&{\cal S}^2 \left(A \otimes B\right) = 
{\cal S}^2 \left(A\right) \otimes {\cal S}^2 \left(B\right)\  \oplus \
\Lambda^2\left(A\right) \otimes \Lambda^2\left(B\right) \nonumber \\
&&{\cal S}^3 \left(A \otimes B\right) = 
{\cal S}^3 \left(A\right) \otimes {\cal S}^3 \left(B\right)\  \oplus \
{\$}^{^{{ {{\hbox{{\tiny {\tiny\renewcommand
\arraycolsep{0.1pt} 
\begin{tabular}{|c|c|}\hline
& \\ \hline
 \\
\cline{1-1}  
\end{tabular}}}}}}}}} \hskip -.2truecm\left(A\right) \otimes 
{\$}^{^{{\tiny \begin{tabular}{|c|c|}\hline
& \\ \hline
 \\
\cline{1-1}  
\end{tabular}}}} \hskip -.2truecm \left(B\right) \ \oplus \
\Lambda^3\left(A\right) \otimes \Lambda^3\left(B\right). \nonumber 
\eeqa

Let $\g_1$ be a representation of $\s \oplus \s$
and let $\D_{1,1}$ be the vector representation of $\s \oplus \s$.
Using the first equation given in \eqref{equiv}, since $\g_1$ is a reducible
representation of $\s \oplus \s$, $\ \Sy^3(\g_1)$ reduces to three types
of terms (i) ${\cal S}^3 \left({\cal} \D\right)$, 
(ii) ${\cal S}^2 \left({\cal} \D\right)  \otimes \D' $
and (iii) ${\cal D} \otimes {\cal D}'\otimes  {\cal D}''$
with ${\cal D}, {\cal D}',{\cal D}''$ three irreducible representations.
Thus all possible $\s \oplus \s-$equivariant mappings are of the type
(i) ${\cal S}^3 \left({\cal} \D\right) \longrightarrow {\cal D}_{1,1}$,
(ii) ${\cal S}^2 \left({\cal} \D\right)  \otimes \D' 
\longrightarrow {\cal D}_{1,1}$  and
(iii) ${\cal D} \otimes {\cal D}'\otimes  {\cal D}''
\longrightarrow {\cal D}_{1,1}$. We now characterise more precisely these
mappings.
\\

\noi
{\bf Explicit description of the $\s \oplus \s$ equivariant mappings}

\medskip
\noindent
(i) \underline{Type $I$. $\s \oplus \s-$ equivariant mappings}:
${\cal S}^3 \left({\cal} \D\right) \longrightarrow {\cal D}_{1,1}$
\\

\noi
Let ${\cal D} = {\cal D}_a \otimes {\cal D}_b$ with $a,b \in
\mathbb N$, 
${\cal D}_{1,1} \subseteq {\cal D}_{a,b}\otimes {\cal D}_{a,b}\otimes 
{\cal D}_{a,b}$ if $a$ and $b$ odd.
>From the third equation of \eqref{equiv}
${\cal D}_{1,1} \subseteq {\cal S}^3\left({\cal D}_{a,b}\right)$ if
either 
\begin{enumerate}
\item[$I_S$ :] ${\cal D}_1 \subseteq {\cal S}^3\left({\cal D}_{a}\right)$ and
${\cal D}_1 \subseteq {\cal S}^3\left({\cal D}_{b}\right)$;
\item[$I_A$ :]  ${\cal D}_1 \subseteq \Lambda^3\left({\cal D}_{a}\right)$ and
${\cal D}_1 \subseteq \Lambda^3\left({\cal D}_{b}\right)$;
\item[$I_M$ :]
${\cal D}_1 \subseteq {\$}^{^{{\tiny \begin{tabular}{|c|c|}\hline
& \\ \hline
 \\
\cline{1-1}  
\end{tabular}}}} \hskip -.2truecm \left({\cal D}_a\right)$ and
${\cal D}_1 \subseteq {\$}^{^{{\tiny \begin{tabular}{|c|c|}\hline
& \\ \hline
 \\
\cline{1-1}  
\end{tabular}}}} \hskip -.2truecm \left({\cal D}_b\right)$.
\end{enumerate} 
We call these  $\s \oplus \s-$ equivariant 
mappings,  mappings of type $I_S$ (symmetric), $I_A$(antisymmetric) 
and $I_M$(mixed) respectively.  
In particular, when   ${\cal D}={\cal D}_{a,a}$ and  $a$ odd,
the mapping ${\cal S}^3\left({\cal D}_{a,a}\right)\longrightarrow
{\cal D}_{1,1}$
is {\it always} $\s \oplus \s-$ equivariant and is called
type $I_0{}_S,I_0{}_A,I_0{}_M$ respectively. 
The extension of the Poincar\'e algebra given in 
Example \ref{FP} is of  type  $I_0{}_M$ with ${\cal D} = {\cal D}_{1,1}$.\\

\noi
(ii) \underline{Type $II$. $\s \oplus \s-$ equivariant mappings}:
 ${\cal S}^2 \left({\cal} \D\right)  \otimes \D' 
\longrightarrow {\cal D}_{1,1}$ 
\\

\noi
Let ${\cal D}= {\cal D}_{a,b}$ and ${\cal D}'= {\cal D}_{c,d}$,
${\cal D}_{1,1} \subseteq {\cal D}_{a,b}\otimes {\cal D}_{a,b}\otimes 
{\cal D}_{c,d}$ if  $c,d$ odd,
and there exists an $n$ such that 
$ 2a-2n-c=1$ or $c-2a + 2n =1$ (and similar relations for $b,d$).
>From the second   equation of \eqref{equiv}
${\cal D}_{1,1} \subseteq {\cal S}^2\left({\cal D}_{a,b}\right)
\otimes {\cal D}_{c,d}$ if either 

\begin{enumerate}
\item[$II_S$ :] ${\cal D}_1 \subseteq {\cal S}^2\left({\cal D}_{a}\right)
\otimes {\cal D}_c$ and
${\cal D}_1 \subseteq {\cal S}^2\left({\cal D}_{b}\right)
\otimes {\cal D}_d$;
\item[$II_A$ :]  ${\cal D}_1 \subseteq \Lambda^2\left({\cal D}_{a}\right)
\otimes {\cal D}_c$ and
${\cal D}_1 \subseteq \Lambda^2\left({\cal D}_{b}\right)
\otimes {\cal D}_d$.
\end{enumerate}
We call these  
 $\s \oplus \s-$ equivariant 
mappings,  mappings of type 
$II_S$ and  $II_A$ 
 respectively. In particular, when $a, b$ even 
${\cal D}_{0,0} \subseteq
{\cal S}^2\left(D_{a,b}\right)$ and
 ${\cal D}_{1,1}
\subseteq {\cal S}^2 \left({\cal D}_{a,b}\right) \otimes {\cal D}_{1,1}$
(type $II_{0S}$).
The extension of the Poincar\'e algebra \eqref{lege-mica} is of type
$II_{0S}$ with $\D'=\D_{2,0}\oplus \D_{0,2}, \D=\D_{1,1}$. \\

\noi
(iii) \underline{Type $III$. $\s \oplus \s-$ equivariant mappings}:
 ${\cal D} \otimes {\cal D}'\otimes  {\cal D}''
\longrightarrow {\cal D}_{1,1}$. 
\\

\noi
Let ${\cal D}= {\cal D}_{a,b}, {\cal D}'= {\cal D}_{c,d}$
and ${\cal D}''= {\cal D}_{e,f}$ with $a \ge c \ge e$, 
${\cal D}_{1,1} \subseteq {\cal D}_{a,b}\otimes {\cal D}_{c,d}\otimes 
{\cal D}_{e,f}$ if  $a+c+e$ and $b+d+f$ are odd
  and if there exists an
$n$ such that $a+c+e -2 n=1$ or $e-a-c+2n=1$
(plus similar relations for $b,d,f$).
There are many $\s \oplus \s-$equivariant mappings of these
types. \\

We now give explicit examples of $\s \oplus \s-$equivariant
mappings of type $I$, $II$ and $III$.\\

\begin{example}
\rm{
 Let $\D=\D_{1}\otimes \D_{1}=\D_{1,0}\otimes \D_{0,1} $.
Using conventional notations for spinors,
let $\D_{1,0} = \left<\psi_\alpha, \alpha=1,2\right>$
and $\D_{0,1}=  \left<\bar \chi^{\dot \alpha}, \dot \alpha=1,2\right>$
be the spinor representations of $\s \oplus \s$.
Introduce the Dirac $\Gamma-$matrices
($\{\Gamma_m,\Gamma_n\}= \Gamma_m \Gamma_n + \Gamma_n \Gamma_m=
2 \eta_{mn} I_4$, with $I_4$ the four dimensional
identity  matrix)

$$\Gamma_m =\begin{pmatrix} 0& \sigma_m \\ \bar \sigma_m&0\end{pmatrix},$$

\noi
where $\sigma_0=\bar \sigma_0$ is the identity matrix
and $\bar \sigma_i=-\sigma_i, i=1,2,3 $ with $\sigma_i$ the Pauli
matrices. The index structure of the $\sigma_m-$matrices
is as follow $\sigma_m \to \sigma_m{}_{\alpha \dot \alpha},
\bar \sigma_m \to \bar \sigma_m{}^{\dot \alpha  \alpha}$.
We also define 
$\psi_\alpha =\varepsilon_{\alpha\beta}\psi^\beta$,
$\psi^\alpha =\varepsilon^{\alpha\beta}\psi_\beta$,
$\bar\chi_{\dot\alpha}=\bar \varepsilon_{\dot\alpha
\dot\beta}\bar\chi^{\dot\beta}$, $\bar\chi^{\dot\alpha}
=\bar \varepsilon^{\dot\alpha\dot\beta}\bar\chi_{\dot\beta}$
with the invariant antisymmetric $\s$ matrices 
$\varepsilon, \bar \varepsilon$  given by
$\varepsilon_{12} = \bar \varepsilon_{\dot 1\dot 2}=-1$,
$\varepsilon^{12} = \bar \varepsilon^{\dot 1\dot 2}=1$.
A direct calculation gives $\bar \sigma_m{}^{\dot \beta \beta}=
\varepsilon^{\beta \alpha} \bar \varepsilon^{\dot \beta \dot \alpha} 
\sigma_m{}_{\alpha \dot \alpha}$. Furthermore since the
Dirac $\Gamma-$matrices are representations of the Clifford
algebra, we have the relations
$\sigma_m \bar \sigma_n +  \sigma_n \bar  \sigma_m = \eta_{mn} \sigma_0,$
and thus $\sigma_m{}_{\alpha \dot \alpha} 
\bar \sigma_n{}^{\dot \alpha\alpha} = \text{Tr} \sigma_m \bar \sigma_n=2 \eta_{mn}.$

Now,  we consider the representation
$$ \D'_{1,0}\cong {\$}^{^{{\tiny \begin{tabular}{|c|c|}\hline
1& 3\\ \hline
2 \\
\cline{1-1}  
\end{tabular}}}} \hskip -.2truecm \left({\cal D}_{1.0}\right).$$

\noi
We introduce the projector  (Young symmetriser)

$$P_{_{{\tiny \begin{tabular}{|c|c|}\hline
1& 3\\ \hline
2 \\
\cline{1-1}  
\end{tabular}}}}= \frac13( 1-(12) +(13) - (123))$$

\noi
with 
$$(a\  b) = \begin{pmatrix} a &b \\
                                b&a\end{pmatrix}, 
(a\   b \ c ) = \begin{pmatrix} a &b & c\\
                                b&c&a\end{pmatrix}$$

\noi two
cycles of ${\cal S}_3$ the group of permutation with three elements.
A direct calculation gives 
$$ P_{_{{\tiny \begin{tabular}{|c|c|}\hline
1& 3\\ \hline 
2 \\ 
\cline{1-1}   
\end{tabular}}}} \Big(\psi_{\alpha} \otimes \psi_{\beta} \otimes 
\psi_{\gamma}\Big)
=\varepsilon_{\alpha \beta} \lambda_\gamma
$$

\noi
and $\D'_{1,0}=\left<\lambda_\alpha, \alpha=1,2\right>$
(the same result can be obtained using the usual calculus of the
Clebsch-Gordan coefficients).
Proceeding along the same lines with $\D_{0,1}$ and
introducing 

$$ \D'_{0,1}\cong {\$}^{^{{\tiny \begin{tabular}{|c|c|}\hline
1& 3\\ \hline
2 \\
\cline{1-1}  
\end{tabular}}}} \hskip -.2truecm \left({\cal D}_{0,1}\right)
=\left<\bar \rho^{\dot \alpha} ,\dot \alpha=1,2\right>$$

\noi
we obtain

$$ 
 P_{_{{\tiny \begin{tabular}{|c|c|}\hline
1& 3\\ \hline 
2 \\ 
\cline{1-1}   
\end{tabular}}}} \Big(\psi_{\alpha} \otimes \psi_{\beta} \otimes 
\psi_{\gamma}\Big)
\bigotimes
 P_{_{{\tiny \begin{tabular}{|c|c|}\hline
1& 3\\ \hline 
2 \\ 
\cline{1-1}   
\end{tabular}}}} \Big(\bar \chi_{\dot \alpha} \otimes 
\bar \chi_{\dot \beta} \otimes 
\bar \chi_{\dot \gamma}\Big)=\varepsilon_{\alpha \beta} \bar 
\varepsilon_{\dot \alpha
\dot \beta} \lambda_\gamma \otimes \bar \rho_{\dot \gamma}.$$

\smallskip
\noi
Symmetrising the R.H.S. we then get

\beqa
\label{fpoincare}
{\cal S}^3\Big((\psi_\alpha \otimes \bar \chi_{\dot \alpha})
\otimes (\psi_\beta \otimes \bar \chi_{\dot \beta}) \otimes
 (\psi_\gamma \otimes \bar \chi_{\dot \gamma})\Big)=
\varepsilon_{\alpha \beta} \bar \varepsilon_{\dot \alpha \dot \beta} 
\lambda_\gamma \bar \rho_{\dot \gamma} +
\varepsilon_{\gamma \alpha } \bar \varepsilon_{\dot \gamma \dot \alpha} 
\lambda_\beta \bar \rho_{\dot \beta} +
\varepsilon_{\beta \gamma} \bar \varepsilon_{\dot \beta \dot \gamma} 
\lambda_\alpha \bar \rho_{\dot \alpha}.
\eeqa

Now, from the isomorphism of $\D_{1,0} \otimes \D_{0,1}$
with the vector representation, 
and using the relation $\sigma_m{}_{\alpha \dot \alpha} \bar
\sigma_n{}^{\dot \alpha  \alpha} = 2 \eta_{mn}$ we have 
the correspondence 

\beqa
\label{spin-vec}
\begin{array}{ll}
V_m =\bar \sigma_m{}^{\dot \alpha \alpha} \psi_\alpha \otimes 
\bar \chi_{\dot \alpha}, & \psi_\alpha \otimes 
\bar \chi_{\dot \alpha} = \frac12 \sigma^m{}_{\alpha \dot \alpha} V_m,\\
P_m =\bar \sigma_m{}^{\dot \alpha \alpha} \lambda_\alpha \otimes 
\bar \rho_{\dot \alpha}, & \lambda_\alpha \otimes 
\bar \rho_{\dot \alpha} = \frac12 \sigma^m{}_{\alpha \dot \alpha} V_m,
\end{array}
\eeqa

\noi
(thus $\left<P_m, \ m=0,\cdots,3\right> \sim \D_{1,1}$,
 $\left<V_m, \ m=0,\cdots,3\right> \sim \D'_{1,1}$)
  and equations \eqref{fpoincare} reduce to 

$$\Sy^3(V_m\otimes V_n \otimes V_p)=
\eta_{mn} P_p + \eta_{np} P_m + \eta_{pm} P_n.$$ 

\noi
We denote $_3 \frak{iso}(1,3,\mathbb C)          =\frak{iso}(1,3,\mathbb C) 
\oplus \D_{1,1}$ 
the corresponding Lie algebra of order 3.
If  we now take  the real form of $\s \oplus \s$ corresponding
to $\mathfrak{sl}(2, \mathbb C)$ (the universal covering group 
of $SO(1,3)$  being $SL(2,\mathbb C)$),  the representation
$\D_{1,0}$ and $\D_{0,1}$ become  complex conjugate. Thus if we take
$\bar \chi_{\dot \alpha} = \left(\psi_\alpha\right)^\star$ (the
complex conjugate of $\psi_\alpha$), and similarly $\bar \rho_{\dot \alpha}
=\left(\lambda_\alpha\right)^\star$, 
$V_m$ and $P_m$ become real vectors of $\mathfrak{so}(1,3).$ 
}\end{example}

\begin{example}
\rm{
 Let $\D=\D_{3,3}$.
Using spinor notations 

$$\D_{3,0} =\left<\psi_{\alpha \beta \gamma},\alpha, \beta, 
\gamma=1,2\right>, \
\D_{0,3} =\left<\bar \chi^{\dot \alpha \dot \beta \dot \gamma}, 
\dot \alpha, \dot \beta, \dot \gamma=1,2\right>,$$ 

\noi
 with $\psi_{\alpha \beta \gamma}, 
\bar \chi^{\dot \alpha \dot \beta \dot \gamma}$ 
symmetric spinor-tensors. This case is more involved than the previous
one, because 

$$ {\$}^{^{{\tiny \begin{tabular}{|c|c|}\hline
1& 3\\ \hline
2 \\
\cline{1-1}  
\end{tabular}}}} \hskip -.2truecm \left({\cal D}_{3,0}\right)$$

\noi
is a reducible representation. However, using the correspondence
\eqref{spin-vec} elements of $\D_{3,3}$ are symmetric traceless 
tensors of order three

$$T_{mnp}= \bar \sigma_m{}^{\alpha \dot \alpha}
\bar \sigma_n{}^{\beta \dot \beta}
\bar \sigma_p{}^{\gamma \dot \gamma}
\psi_{\alpha \beta \gamma} 
\bar \chi_{\dot \alpha \dot \beta \dot\gamma}$$

\noi
(the symmetry of $T$ comes from the symmetry of $\psi$ and $\bar \chi$ and
$T_{mnp} \eta^{nm}=0$ from $\sigma_{m}{}_{\alpha \dot \alpha} 
\sigma^m{}_{\beta \dot \beta}= 2\varepsilon_{\alpha \beta} 
\bar  \varepsilon_{\dot \alpha \dot \beta}$). Now it is easy to
see that the mapping 

$$ T_{m_1 n_1 p_1}  \otimes  T_{m_2 n_2 p_2} \otimes T_{m_3 n_3 p_3}
\longrightarrow \eta_{m_1 m_2} \eta_{n_1 n_2} \eta_{p_1 m_3}
\eta_{p_2 n_3} P_{p_3}
$$

\noi is $\s \oplus \s-$equivariant. Thus, symmetrising the R.H.S we have

\beqa
 &\Sy^3\big(T_{m_1 n_1 p_1}  \otimes  T_{m_2 n_2 p_2} \otimes T_{m_3 n_3 p_3}
\big) =\\
&\eta_{m_1 m_2} \eta_{n_1 n_2} \eta_{p_1 m_3} \eta_{p_2 n_3} P_{p_3}+
\eta_{m_2 m_3} \eta_{n_2 n_3} \eta_{p_2 m_1} \eta_{p_3 n_1} P_{p_1}+
\eta_{m_3 m_1} \eta_{n_3 n_1} \eta_{p_3 m_2} \eta_{p_1 n_2} P_{p_2}.
\nonumber 
\eeqa

}
\end{example}

\begin{example}
\rm{
 Let $\D=\D_{1,1}\oplus \D_{2,0} \oplus \D_{0,2}$, we
have $\mathbb C \subseteq \Sy^2(\D_{2,0}), \ 
\mathbb C \subseteq \Sy^2(\D_{0,2})$.
Let $\D_{2,0}\oplus \D_{0,2}= \left<A_{i}, i=1,2,3\right>
\oplus \left<\bar A_{i}, i=1,2,3\right>$.
 From 
 $\Sy^2(A_i \otimes A_j) = \delta_{ij}$ and 
$\Sy^2(\bar A_i \otimes \bar A_j) = \delta_{ij}$ the mappings
$\Sy^3\Big(\D_{1,1} \oplus \D_{2,0} \oplus \D_{0,2}\Big) \supseteq 
\Big(\Sy^2(\D_{2,0})  \oplus \Sy^2(\D_{0,2}) \Big)\otimes
\D_{1,1} \longrightarrow \D_{1,1}$ follows immediately. 
This gives  the trilinear brackets obtained in \eqref{lege-mica}.\\
}
\end{example}

\begin{example}
\rm{
 Let $\D=\D_{1,0} \oplus \D_{0,1} \oplus
\D_{0,0}$. The mapping $\Sy^3(\D_{1,0} \oplus \D_{0,1} \oplus
\D_{0,0}) \supset \D_{1,0} \oplus \D_{0,1} \oplus
\D_{0,0} \longrightarrow \D_{1,1}$ is immediate.
Such a mapping was considered in \cite{Michel-finit}.
}
\end{example}

\subsection{Lie algebras of order $3$ associated to the Poincar\'e algebra}
Let us denote $\V$ the vector space isomorphic to
$\D_{1,1}$ generated by the vectors $P_m, m=0,\cdots,3$. 
Now, from the characterisation of $\s \oplus \s-$equivariant mappings 
from  $\Sy^3(\D) $ into $ \D_{1,1}$ and lemma \ref{rep}, we construct
Lie algebras of order $3$ whose  zero graded part is isomorphic to the
Poincar\'e algebra.

\begin{theorem}
\label{Fpoincare}
Let $\D$ be a reducible representation of $\s \oplus \s$ such that:
\begin{enumerate}
\item  the action of $\s \oplus \s$ on ${\cal D}$ extends to an
action of $\i$ on ${\cal D}$  as in Lemma \ref{rep};
\item there exist an $\s \oplus \s-$equivariant mapping from
$\Sy^3(\D)\longrightarrow \V$.
\end{enumerate}
Then if $\g= \i \oplus \D$ is a Lie algebra of
order $3 $, the action 
of $\V$ on $\D$ is trivial.
\end{theorem}

\noi
{\it Proof.} \\
Type $I$ : Assume $\Sy^3(\D_{a,b}) \longrightarrow \V$.
This means that $a,b$ are odd.
\begin{itemize}
\item[1.] Suppose there exists a representation $\D' \subseteq \D$
(not necessarily irreducible) such that$\  \D'  \subseteq [\V, \D_{a,b}]$.
The Jacobi identity J4 with $Y_1=Y_2=Y_3=Y_4 \in \D_{a,b}$ leads
to a contradiction, thus $[\V, \D_{a,b}]=0$.

\item[2.]  Suppose there exist a representation $\D_{c,d} \subseteq \D$
such that  $ \D_{a,b} \subseteq  [\V, \D_{c,d}] $.
Since $a,b$ are odd, $c,d$ are even and 
thus  $\V \subseteq \hskip -.3truecm \slash \ \Sy^2(\D_{a,b}) \otimes D_{c,d}$.
The Jacobi identity J4 with $Y_1, Y_2, Y_3 \in \D_{a,b}, Y'_4 \in \D_{c,d}$\
together with $\Sy^2(\D_{a,b}) \otimes \D_{c,d}=0$
gives a contradiction and $[\V,\D_{c,d}]=0$.
\end{itemize}

\noi
Type $II : $ Assume $\V \subseteq \Sy^2(\D_{a,b}) \otimes \D_{c,d}$
and $\Sy^3(\D_{a,b}) = \Sy^3(\D_{c,d})=0$ (not of type $I$). In this
case $c,d$ are odd.

\begin{itemize}
\item[1.]  Suppose $ \D_{c,d} \subseteq[\V, \D_{a,b}]$.
The Jacobi identity J4 with $ Y_1= Y_2=Y_3  \in \D_{a,b},Y'_4 \in \D_{c,d}$
gives a contradiction and $[\V, \D_{a,b}] =0$.
\item[2.] Suppose $\D_{a,b} \subseteq[\V, \D_{c,d}]  $, thus $a,b$ are even
and $\V \nse  \Sy^2(\D_{c,d})\otimes \D_{a,b}$.
The Jacobi identity J4 with $ Y_1=Y_2 \in \D_{a,b}, Y_3'=Y_4'\in \D_{c,d}$
gives a contradiction and $[\V, \D_{a,b}] =0$.
\item[3.] Suppose there exists $\D_{e,f} \subseteq \D$, with 
$\D_{e,f} \ne \D_{c,d}$, such that $\D_{e,f} \subseteq [\V,\D_{a,b}]$. 
The same argument as in the point 1. above
gives $[\V, \D_{a,b}]=0$.
\item[4.]  Suppose there exists $\D_{e,f} \subseteq \D$, with 
$\D_{e,f} \ne \D_{a,b}$, such that $ \D_{e,f} \subseteq[\V,D_{c,d}] $. 
The Jacobi identity J4 with $Y_1=Y_2=Y \in \D_{ab}$ and 
$Y_3'=Y_4'=Y' \in \D_{cd}$ gives
$[Y,\{Y,Y',Y'\}] + [Y',\{Y,Y,Y'\}] =0.$ If we suppose  that 
$\{Y,Y',Y'\}= P \in \V$ we know from the points 1. and 3. above
that $[P,Y]=0$. Thus the previous identity becomes $[Y',\{Y,Y,Y'\}]=
[Y',P]=0$
and $[\V,D_{c,d}] =0.$
\item[5.] Suppose there exists $\D_{e,f} \subseteq \D$, with 
$\D_{e,f} \ne \D_{a,b}$ and $ \D_{e,f}\ne \D_{c,d}$ such that 
either $\D_{a,b} \subseteq [\V,\D_{e,f}]$ or 
$\D_{c,d} \subseteq [\V,\D_{e,f}]$. The Jacobi identity J4 with
$Y_1=Y_2=Y \in \D_{a,b}, Y' \in \D_{c,d}, Y'' \in \D_{e,f}$ gives
$2[Y,\{Y,Y',Y''\}]+ [Y',\{Y,Y,Y''\}]+ [Y'',\{Y,Y,Y'\}]=0$.
If $\{Y,Y',Y''\} \in \V$ or $\{Y,Y,Y''\} \in \V$ since  
$[\V, \D_{a,b}]=[\V,\D_{c,d}]=0$
(see 1.,2.,3. and 4. above), the previous identity reduces to 
$[Y'',\{Y,Y,Y'\}]=[Y'',P]=0$ and thus $[\V,\D_{e,f}]=0$. 
\end{itemize}

\noi
Type $III : $ Assume $\V \subseteq \D_{a,b} \otimes \D_{c,d} \otimes \D_{e,f}$,
$\Sy^3(\D_{a,b}) = \Sy^3(\D_{c,d})=\Sy^3(\D_{e,f})=
0$ (not of type $I$) and $\Sy^2(\D_{a,b}) \otimes \D_{c,d}=
\Sy^2(\D_{a,b}) \otimes \D_{e,f}=\Sy^2(\D_{c,d}) \otimes \D_{a,b}=
\Sy^2(\D_{c,d}) \otimes \D_{e,f}=\Sy^2(\D_{e,f}) \otimes \D_{a,b}=
\Sy^2(\D_{e,f}) \otimes \D_{c,d}=0$ (not of type $II$).
\begin{itemize}
\item[1.] If we assume $[\V,\D_{a,b}] \subseteq \D_{g,h}$ with
$\D_{g,h} = \D_{c,d}$ or $\D_{g,h} \ne \D_{c,d}, \D_{g,h} \ne \D_{e,f}$,
the Jacobi identity J4 with $Y_1, Y_2 \in \D_{a,b}, Y'_3 \in \D_{c,d},
Y''_4 \in \D_{e,f}$ leads to a contradiction and $[\V, \D_{a,b}]=0$.
\item[2.] Suppose there exists $\D_{g,h}$ such that 
$[\V,\D_{g,h}] \ \nse \ \D_{a,b}$, the Jacobi identity J4 with 
$Y_1 \in \D_{a,b}, Y_2 \in \D_{c,d}, Y_3 \in \D_{e,f}$ and 
$ Y_4 \in D_{g,h}$ leads to a
contradiction and thus $[\V,\D_{g,h}]=0.$
\end{itemize}
This means that the action of $\V$ on $\D$ is trivial and
thus $[\V,\D]=0$. The remaining Jacobi identities are easy to be checked.
Which ends the proof. QED.

\begin{corollary}
With the hypothesis of theorem \ref{Fpoincare} the action of $\V$ on $\D$ is trivial
{\it i.e.} $[\V,\D]=0$.
\end{corollary}

\begin{remark}
\label{general}
Differently as in theorem \ref{Fpoincare} if $\g=(\s \oplus \s \oplus \V)\oplus \D$ is a Lie
algebra of order 3 satisfying $[\V,\D]=0$ then $\Sy^3(\D) \longrightarrow \V.$
Indeed  if  we suppose for contradiction that  
$\Sy^3(\D) \longrightarrow \s \oplus \s$, and let $Y_1, Y_2,Y_3 \in \D$
such that
$\{Y_1,Y_2,Y_3\} =a  L \text{ with } L\in \s \oplus \s, a \in \mathbb C $.
The Jacobi identity J3 with $P \in \V$ and $Y_1,Y_2,Y_3$ as above leads to
$a=0$ since the elements of $\D$ commute with the
elements of $\V$ and $L$ do not commute with $P$.
 \end{remark}

The following property classify all Lie algebras of order 3 based on the Poincar\'e algebra
such that the representation $\D$ is of dimension 4.

\begin{proposition}
Let $\g=\i\oplus \D$ be an elementary 
Lie algebra of 
order 3, with  $\D$ a representation 
of dimension $4$. Then, 
\begin{enumerate}
\item $\D \cong \D_{1,1}$;

\item $[\V,\D_{1,1}] =0$;
\item $\g\cong \  _3\i$ (the  complexified of the  
Lie algebra of order three of Example \ref{FP}).
\end{enumerate}
\end{proposition}

\noi
{\it Proof.}
1. Since the representation $\D_\ell, \ell \in \mathbb N$
 of $\s$ is of dimension
$\ell +1$, 
the four dimensional representations of $\s \oplus \s$ are (up a permutation of the action 
$\s \oplus \s$) :

$$
\D_{3,0}, \  \D_{2,0} \oplus \D_{0,0},\ 
\D_{1,1}, \ \D_{1,0}\oplus \D_{0,1}, \ 
\D_{1,0} \oplus \D_{0,0} \oplus \D_{0,0},\ 
\D_{0,0} \oplus \D_{0,0} \oplus \D_{0,0} \oplus \D_{0,0}.
$$

\noi
Since $ \Sy^3(\D) \longrightarrow \i$,  
a simple weights argument shows that the only possibilities are 
(i) $\D=\D_{1,1}$ with $\V \subseteq \Sy^3(\D)$ 
(in agreement with the explicit description of $\s \oplus \s-$equivariant
mappings)
or (ii) by Example 2.4
$\s \cong \D_{2,0}  \subseteq \Sy^3(\D_{2,0})$. In the
second case we have by Remark  \ref{irred} $[\V,\D_{2,0}]=0$ and
thus the Jacobi identity J3 with $P \in \V, Y_1, Y_2, Y_3 \in \D_{2,0}$
leads to a contradiction
since $P$ commute with $Y_i$ and not commute with $U$
 (see Eq.\eqref{poincare}). Thus,  the only non-trivial Lie 
algebra of order 3 is then constructed with $\D_{1,1}$.

2. Since $\D$ is an irreducible representation of $\s \oplus \s$,
by Remark   \ref{irred}, the action of $\V$ on $\D$ is trivial and
$[\V,\D]=0.$

3. Since $[\V, \D] =0$, by Corollary \ref{general}, we have
$ \V \subset \Sy^3(\D)$ (a simple weights argument, as we have seen in
1. above,
also  show that $\s \oplus \s \nse \Sy^3(\D)$). This means, introducing 
$v_{++},v_{+-},v_{-+},v_{--}$ a basis of $\D$
(with notations similar to \eqref{vect}), and using a simple weights argument, 
that the only non-trivial trilinear brackets are:

\beqa
\label{trilin}
\begin{array}{ll}
\left\{v_{++}, v_{++},v_{--}\right\}= \alpha_1 P_{++},&
\left\{v_{--}, v_{--},v_{++}\right\}= \beta_1 P_{--},\nonumber  \\
\left\{v_{++}, v_{+-},v_{-+}\right\}= \alpha_2 P_{++},&
\left\{v_{--}, v_{-+},v_{+-}\right\}= \beta_2 P_{--}, \\
\left\{v_{++}, v_{+-},v_{--}\right\}= \gamma_1 P_{+-},&
\left\{v_{--}, v_{-+},v_{++}\right\}= \delta_1 P_{-+},  \\
\left\{v_{+-}, v_{+-},v_{-+}\right\}= \gamma_2 P_{+-},&
\left\{v_{-+}, v_{-+},v_{+-}\right\}= \delta_3 P_{-+}.  \nonumber
\end{array}
\eeqa

\noi
The action of $\s \oplus \s$ on $\D$ is given by (see \eqref{poincare})

\beqa
\label{bilin}
\begin{array}{ll}
\left[U_+, v_{- \varepsilon}\right]=v_{+\varepsilon},
&\left[V_+, v_{ \varepsilon-}\right]=-v_{\varepsilon +}, \\
\left[U_-, v_{+\varepsilon }\right]=v_{-\varepsilon},&
\left[V_-, v_{ \varepsilon +}\right]=-v_{\varepsilon -}, \\
\left[U_0, v_{\varepsilon \varepsilon'}\right]=
\varepsilon v_{\varepsilon \varepsilon'},&
\left[V_0, v_{\varepsilon \varepsilon'}\right]=
\varepsilon' v_{\varepsilon \varepsilon'}, 
\end{array}\nonumber
\eeqa

\noi
with $\varepsilon, \varepsilon'=\pm$.
The Jacobi identity J3 gives

$$\alpha_2 = -\frac12 \alpha_1, \beta_1 = \alpha_1,
\beta_2 =-\frac12 \alpha_1, \gamma_1=\frac12  \alpha_1,
\gamma_2 = - \alpha_1,
\delta_1=\frac12  \alpha_1,\delta_2 = - \alpha_1.$$

\noi
Il $\alpha_1\ne 0$,  we set
\beqa
\begin{array}{ll}
v_0=-\sqrt[3]{\frac{1}{2 \alpha_1}}\left(v_{+-}+v_{-+}\right),&
v_3=-\sqrt[3]{\frac{1}{2 \alpha_1}}\left(v_{+-}-v_{-+}\right),
 \\
v_1=-\sqrt[3]{\frac{1}{2 \alpha_1}}\left(v_{++}+v_{--}\right),&
v_2=i \sqrt[3]{\frac{1}{2 \alpha_1}}\left(v_{++}-v_{--}\right),
\end{array}
\nonumber
\eeqa

\noi
 and we get
$$\left\{v_\mu, v_\nu,v_\rho\right\}=
\eta_{\mu \nu} P_\rho +
\eta_{\mu \rho} P_\nu + \eta_{\nu \rho} P_\mu.$$

\noi So, the algebra  is isomorphic to the complexified 
 elementary Lie algebra of order 3 of Example 2.5. 
This results remains true if we consider its real  form
 corresponding to the Lie algebra of order 3
of example 2.5. 
In this case if $\alpha_1 >0$ we rescale the coefficients by 
$-\sqrt[3]{\frac{1}{2 \alpha_1}}$ as above, and when  $\alpha_1 <0$
we rescale the coefficients by $\sqrt[3]{\frac{-1}{2 \alpha_1}}$.
Q.E.D.

\section{Contractions of the Poincar\'e-algebra of order $3$}
\renewcommand{\theequation}{5.\arabic{equation}}   
\setcounter{equation}{0}
 
\subsection{Contractions of elementary Lie algebras of order $3$}
\label{3-contractions}

The variety ${\cal F}_{m,n}$ being an algebraic variety, one can naturally 
endow it with the Zariski topology.

\begin{definition}
A contraction of $\p$ is a point $\p'\in {\cal F}_{m,n}$ such that $\p'\in \overline{ {\cal O}}_\p$, the closure in the Zariski sense.
\end{definition}

In the complex case, the notion of contraction is equivalent to the following.
Let $\p=(\p_1,\p_2,\p_3)$ be a given multiplication of elementary Lie algebras of order $3$, $\g=\g_0\oplus\g_1$ 
and let $(h_p)_{p\in\NN}$ (with $h_p=(h_{0,p},h_{1,p})\in GL({m,n})$) be a sequence of isomorphisms. Define
$\p_{p}=(\p_{1,p},\p_{2,p},\p_{3,p})$ 
by
\beqa
\label{fiurile}
\p_{1,p} (X_1, X_2)&=&h_{0,p}^{-1} \p_1 (h_{0,p} (X_1), h_{0,p} (X_2)), \nonumber \\
\p_{2,p} (X_1, Y_2)&=&h_{1,p}^{-1} \p_2 (h_{0,p} (X_1), h_{1,p} (Y_2)),  \\
\p_{3,p} (Y_1, Y_2, Y_3) &=& h_{0,p}^{-1} \p_3 ( h_{1,p} (Y_1), h_{1,p} (Y_2), h_{1,p} (Y_3))\ .\nonumber 
\eeqa
\noi
If 
the limit  
$
\label{limita} \displaystyle{\lim_{p \to  + \infty}} \p_p
$
exists,
this limit is in the closure of  
$\mathcal{O}_{\varphi}$. Then it
is a contraction of $\varphi$. Note that, in the complex case, every  
contraction of $\varphi$
is obtained by this way \cite{Go2}.

\bigskip

Moreover, In\"on\"u-Wigner contractions \cite{inonu-wigner} turn out to be a relevant subclass of contractions. We consider the
automorphisms $h_\e=(h_{0,\e}, h_{1,\e})$ of the form $h_{0,\e}=h_{0}^{(1)} 
+ \e h_0^{(2)}$ and 
$h_{1,\e}=h_{1}^{(1)} + \e h_1^{(2)}$ with $h_0^{(1)}, h_1^{(1)}$ singular, 
$h_0^{(2)}, h_1^{(2)}$ regular
 and $\e$ infinitesimal.
A further particularisation 
inspired from the Weimar-Woods construction \cite{ww-2} is given by
$$h_\e={\rm diag} (\e^{a_1},\dots,\e^{a_m},\e^{b_1},\dots, \e^{b_n})$$
with $a_i,b_j\in\ZZ$ ($i=1,\dots,m,\ j=1,\dots,n$). Hence 
$h_{0,\e} (X_i)=\e^{a_i} X_i$ and  $h_{0,\e} (Y_j)=\e^{a_j} Y_j$
and 
 \eqref{fiurile} become
\beqa
\p_{1,\e} (X_i, X_j)=\e^{a_i+a_j-a_k}C_{ij}^k X_k,\nonumber\\
\p_{2,\e} (X_i, Y_j)=\e^{a_i+b_j-b_k}D_{ij}^k Y_k,\nonumber\\
\p_{3,\e} (Y_i, Y_j,Y_k)=\e^{b_i+b_j+b_k-a_\ell}E_{ijk}^\ell X_\ell.
\eeqa
\noi
As already stated, one can define a contraction if the limit $\e\to 0$  
exists, {\it i.e.}
 if $a_i+a_j-a_k\ge 0$, $a_i+b_j-b_k\ge 0$ and $b_i+b_j+b_k-a_\ell\ge 0$ 
for any $a$ and $b$.   

\medskip

\noindent {\bf Examples. }
\label{ex-contractie}

Let  ${\cal F}_{1,1}$ be the algebraic variety of $2=(1+1)-$dimensional
 elementary Lie algebras of order $3$. We consider a 
basis $\{X,Y\}$ of $\g=\g_0\oplus\g_1$ adapted for this decomposition.

\begin{proposition}
\label{dim3-ele}
Any  two-dimensional Lie algebra of order $3$ $\g=\g_0\oplus\g_1$ 
is isomorphic to one of the following Lie algebras of order $3$
\begin{enumerate}
\item $\g^3_1$: $\{Y,Y,Y\}=X,\ [X,Y]=0$;
\item $\g^3_2$: $[X,Y]=Y, \{Y,Y,Y\}=0$;
\item  $\g^3_3$: the trivial Lie algebra of order $3$.
\end{enumerate}
\end{proposition}
\textit{Proof.} We consider the most general possibility for the structure constants of a two-dimensional elementary Lie algebra of order $3$:
\beqa
\label{genu-ele}
\pard X,Y \pari= \alpha_1 Y, 
 \{Y,Y,Y\}=\alpha_2 X\ .
\eeqa
The Jacobi identities {J1-J4} imply $\alpha_1\alpha_2=0$, and we   obtain
\beqa
\label{sol3-ele}
&&\g^3_1,  \alpha_1=0, \alpha_2=1; \nonumber \\
&&\g^3_2,  \alpha_1=1, \alpha_2=0; \nonumber \\
&&\g^3_3,  \alpha_1=0, \alpha_2=0. \nonumber
\eeqa
QED

\begin{corollary}
\label{corolar-ele}
The variety ${\cal F}_{1,1}$ of $2-$dimensional elementary  Lie algebras of order $3$, 
 is the union of two irreducible algebraic components $U_1$ and $U_2$ with
$$ U_1=\overline{\cal O}_{{\g}^3_1} \mbox{ and }  U_2=\overline{\cal O}_{{\g}^3_2}.$$
\end{corollary}
\textit{Proof.} 
One has the following contraction scheme

$$
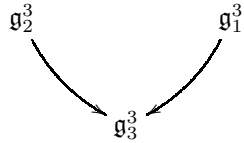
\xymatrix{
 \g^3_2 \ar@/_/[dr] & & \g^3_1 \ar@/^/[dl]\\
& \g^3_3 & }
$$
where by $A\to B$ we denote a contraction of the algebra $A$ to the algebra $B$ ($B$ is a contraction of $A$). QED

\begin{remark}
The algebra $\g_1^3$ has been considered in \cite{FSUSY4,FSUSY6,FSUSY2,FSUSY5,FSUSY,FSUSY3}.
\end{remark}

\subsection{Contraction which leads to a non-trivial extension of the Poincar\'e algebra}
\label{another}

Let $\g = {\mathfrak {so}}(2,3) \oplus {\rm ad}\, {\mathfrak {so}}(2,3)$.
Using vector indices of $\mathfrak{so}(1,3)$ coming from the
inclusion $\mathfrak{so}(1,3)\subset  \mathfrak{so}(2,3)$  we introduce
$\{M_{mn}=-M_{nm}, M_{m4}=-M_{4m}, m,n=0,\dots,3, \ m<n\}$   a basis of ${\mathfrak {so}}(2,3)$ and  
$\{J_{mn}= -J_{nm}, J_{m4}=-J_{4m},  m,n=0,\dots,3, \ m<n\}$  the corresponding basis of 
${\rm ad}\, {\mathfrak {so}}(2,3)$. 
 The multiplication law $\p$ 
of the elementary Lie algebra of order $3$ ${\mathfrak {so}}(2,3) \oplus {\rm ad}\, {\mathfrak {so}}(2,3)$ writes

\beqa
\label{p-so23}
\p_1(M_{mn}, M_{pq})&=&-\eta_{nq} M_{mp}-\eta_{mp} M_{nq}+\eta_{mq} M_{np}+\eta_{np} M_{mq},\nonumber\\
\p_1(M_{mn}, M_{p4})&=&-\eta_{mp} M_{n4} +\eta_{np} M_{m4},\nonumber\\
\p_1 (M_{m4}, M_{p4})&=&-M_{mp},\nonumber\\
\p_2(M_{mn}, J_{pq})&=&-\eta_{nq} J_{mp}-\eta_{mp} J_{nq}+\eta_{mq} J_{np}+\eta_{np} J_{mq},\nonumber\\
\p_2(M_{mn}, J_{p4})&=&-\eta_{mp} J_{n4} +\eta_{np} J_{m4},\\
\p_2(M_{m4}, J_{pq})&=&-\eta_{mp} J_{4q} +\eta_{mq} J_{4p},\nonumber\\
\p_2 (M_{m4}, J_{p4})&=&-J_{mp},\nonumber\\
\p_3 (J_{mn},J_{pq}, J_{rs})&=&(\eta_{mp}\eta_{nq}-\eta_{mq}\eta_{np}) M_{rs}+
 (\eta_{mr}\eta_{ns}-\eta_{ms}\eta_{nr})M_{pq}+(\eta_{pr}\eta_{qs}-\eta_{ps}\eta_{qr}) M_{mn},\nonumber\\
\p_3 (J_{mn},J_{pq}, J_{r4})&=&(\eta_{mp}\eta_{nq}-\eta_{mq}\eta_{np}) J_{r4},\nonumber\\
\p_3 (J_{mn},J_{p4}, J_{r4})&=&\eta_{pr} M_{mn},\nonumber\\
\p_3 (J_{m4},J_{p4}, J_{r4})&=&\eta_{mp} M_{r4}+\eta_{mr} M_{p4}+\eta_{pr} M_{m4},\nonumber
\eeqa
\noi
where $\eta_{mn}= \text{diag}(1,-1,-1,-1)$.
In \cite{Michel-finit},  an In\"on\"u-Wigner contraction was done  
\beqa
\label{F-contractia}
L_{mn}=h_0 (M_{mn})= M_{mn},\nonumber \\
P_m= h_0 (M_{m4})=\e M_{m4}, \nonumber\\
V_{mn}=h_1 (J_{mn})= \sqrt[3]{\e} J_{mn},\nonumber\\
V_m =h_1 (J_{m4})=  \sqrt[3]{\e}  J_{m4},
\eeqa
\noi
and the limit $\e \to 0$ realised the contraction.  
The contracted Lie algebra of order $3$ 
is given by 
\beqa
\label{lege-mica}
\p_1(L_{mn}, L_{pq})&=&-\eta_{nq} L_{mp}-\eta_{mp} L_{nq}+\eta_{mq} L_{np}+\eta_{np} L_{mq},\nonumber\\
\p_1(L_{mn}, P_{p})&=&-\eta_{mp} P_{n} +\eta_{np} P_{m},\nonumber\\
\p_1 (P_m, P_p)&=&0,\nonumber\\
\p_2(L_{mn}, V_{pq})&=&-\eta_{nq} V_{mp}-\eta_{mp} V_{nq}+\eta_{mq} V_{np}+\eta_{np} V_{mq},\nonumber\\
\p_2(L_{mn}, V_{p})&=&-\eta_{mp} V_{n} +\eta_{np} V_{m},\nonumber\\
\p_2(P_{m}, V_{pq})&=&0,\nonumber\\
\p_2(P_{m}, V_{p})&=&0,\nonumber\\
\p_3 (V_{mn},V_{pq}, V_{rs})&=&0,\nonumber\\
\p_3 (V_{mn},V_{pq}, V_{r})&=&(\eta_{mp}\eta_{nq}-\eta_{mq}\eta_{np}) P_{r},\nonumber\\
\p_3 (V_{mn},V_{p}, V_{r})&=&0,\nonumber\\
\p_3 (V_{m},V_{p}, V_{r})&=&\eta_{mp} P_{r}+\eta_{mr} P_{p}+\eta_{pr} P_{m}.
\eeqa
\noi
Thus $L_{mn},P_m$ generate the Poincar\'e algebra and 
$V_{mn}$ and $V_m$ are in the adjoint and vector representations of ${\mathfrak {so}}(1,3)$. 

\begin{remark}
The subalgebra generated by $L_{mn}, P_m$ and $V_{m}$ is the  
 algebra of Example \ref{FP}.
\end{remark}

\section{Deformations of Lie algebras of order $3$}
\renewcommand{\theequation}{6.\arabic{equation}}   
\setcounter{equation}{0}
\subsection{Definition}

Let $\g=\g_0\oplus \g_1$ be a Lie algebra of order $3$ on the complex field (or on a field of characteristic zero).
Let $A$ be a commutative local $\K$-algebra, $\m$ its maximal ideal. We assume that the residual field is isomorphic to $\K$. 
A $A$-Lie  algebra of order $F$ is a Lie algebra of order $F$ whose coefficients in a $\K$-basis belongs to $A$.
We will assume that $A$  admits an augmentation 
 $\epsilon: A \to \K$.
The ideal $m_\epsilon:= \text{Ker\,} \epsilon$ is 
the maximal ideal of $A$. 
 In this context, a deformation $\lambda$ of $\g$ with the
base $(A,m)$ or simply with the base $A$, is a $A$-Lie algebra of order $F$
on the tensor product $A\otimes_\K\K^n$ with brackets 
$[.,.]_\lambda$ and $\{.,.,\}_\lambda $ 
satisfying the Jacobi identities J1-J4.
Sometimes, we add hypothesis on $A$, for example $A$ is finitely generated or $A$ 
is noetherian. In this paper, following 
\cite{magnificul2},
we assume that $A$ is a valuation algebra. It is a local algebra with the following 
property : if $x$ belongs to the field of fractions of $A$ but not
to $A$, then the converse $x^{-1}$ belongs to $\m$. The most interesting examples of 
deformations (the formal deformations of Gerstenhaber, 
the nonstandard perturbations) satisfy this hypothesis. Moreover, a deformation in a 
valued algebra always admits 
a finite decomposition, that is is defined by a finite number of ideals of $A$. Then 
every deformation has a finite Krull dimension
and this means that an hypothesis such as the noetherian property is superfluous. 

\subsection{The Gerstenhaber products}
The Gerstehaber products have been introduced in \cite{Ge1} and  
\cite{Ge2}  to study the deformations of
associative algebras and the structure of Hochschild cohomology
groups. 
Let $\p=(\p_1,\p_2,\p_3)$ be a given multiplication of elementary Lie algebras of 
order $3$, $\g=\g_0\oplus\g_1$ 
The identities { J1-J4} are equivalent to
\beqa
\label{defo-Jacobi}
&&\varphi_1 (\p_1(X_1, X_2), X_3)+\varphi_1 (\p_1(X_3, X_1), X_2)+
\varphi_1 (\p_1(X_2, X_3), X_1)=0,\nonumber \\
&&\varphi_2 (\varphi_1 (X_1, X_2), Y)+\p_2 (\p_2 (X_2, Y), X_1) + \p_2 
(\p_2 (Y, X_1), X_2) = 0,  \\
&& \p_1 (X, \p_3 (Y_1, Y_2, Y_3)) - \p_3 (\p_2 ( X, Y_1), Y_2, Y_3) - 
\p_3 (Y_1, \p_2 (X, Y_2), Y_3) - \p_3 (Y_1,Y_2,\p_2(X,Y_3)) = 0,\nonumber \\
&& \p_2 (Y_1, \p_3 (Y_2, Y_3, Y_4)) + \p_2 (Y_2, \p_3 (Y_1, Y_3, Y_4))+
\p_2 (Y_3, \p_3 (Y_1, Y_2, Y_4))+\p_2 (Y_4, \p_3 (Y_1, Y_2, Y_3))=0.\nonumber
\eeqa
\noi
If  $\p$ and $\p'$ are two products of elementary Lie algebras of order $3$, we 
can define

\beqa
\label{def-criptic}
\begin{array}{llll}
\p \circ_1 \p' :& (\g_0\otimes \g_0\otimes\g_0)&\to& \g_0 \\
 &X_1 \otimes X_2 \otimes X_3 &\mapsto &\varphi_1 (\p'_1(X_1, X_2), X_3)+\varphi_1 
(\p'_1(X_3, X_1), X_2)+\varphi_1 (\p'_1(X_2, X_3), X_1),\\ \\
\p \circ_2 \p' :& (\g_0\otimes \g_0\otimes\g_1)& \to&  \g_1 \\
&X_1\otimes X_2 \otimes Y &\mapsto& \varphi_2 (\varphi'_1 (X_1, X_2), Y)+\p_2 (\p'_2 (X_2, Y), X_1) + \p_2 (\p'_2 (Y, X_1), X_2),\\ \\
\p \circ_3 \p':& (\g_0\otimes {\cal S}^3 (\g_1))&\to& \g_0 \\
& X\otimes (Y_1,Y_2,Y_3)&\mapsto &\p_1 (X, \p'_3 (Y_1, Y_2, Y_3)) - \p_3 (\p'_2 ( X, Y_1), Y_2, Y_3)  \\
&&& - \p_3 (Y_1, \p'_2 (X, Y_2), Y_3)-\p_3 (Y_1,Y_2,\p'_2(X,Y_3)),\\ \\
\p \circ_4 \p' :& (\g_1\otimes {\cal S}^3 (\g_1))&\to& \g_1  \\
&Y_1\otimes (Y_2,Y_3,Y_4)&\mapsto &\p_2 (Y_1, \p'_3 (Y_2, Y_3, Y_4)) + \p_2 (Y_2, \p'_3 (Y_1, Y_3, Y_4)) \\
&&&
+\p_2 (Y_3, \p'_3 (Y_1, Y_2, Y_4)) +\p_2 (Y_4, \p'_3 (Y_1, Y_2, Y_3)).
\end{array}
\eeqa

\begin{proposition}
The map $\p$ endows $\g$ with a structure of  elementary Lie algebra of order $3$ iff
\beqa
\label{cond-defo}
\p \circ_i \p =0 \mbox{ for }i=1,\dots,4.
\eeqa
\end{proposition}

\subsection{Gerstenhaber deformations}
In this section we assume that the  algebra of valuation $A$ is the algebra $\CC[[t]]$ of
formal sequence.
Its law is written  $\p_t$ 
\beqa
\label{def-defo}
\p_t : A \to (\g_0 \oplus \g_1)\otimes  {\CC}[[t]]
\eeqa
\noi
with
\beqa
\label{defo-gen}
\p_t=\p + t \psi^{(1)} + t^2 \psi^{(2)}+\dots +t^n \psi^{(n)} + \cdots, 
\eeqa
\noi 
where the $\psi^{(i)}$'s are linear applications from $A$ to $\g$,  satisfying \eqref{cond-defo}.

\begin{proposition}
\label{defo-general}
Considering a deformation $\p_t$ of $\p$, the maps $\psi^{(p)}$ (with $p\in\NN$) satisfy the equations
\beqa
\label{sol-gen}
\sum_{p+q=r} \psi^{(p)} \circ_i \psi^{(r)}=0, \mbox{ for any }i=1,\dots,4,\ r\in\NN
\eeqa
\noi
where  $\psi^{(0)}=\p$.
\end{proposition}
{\it Proof.} As $\p_t$ is a deformation of $\p$ it satisfies $\p_t \circ_i \p_t=0$. 
For $i=1$, equation \eqref{sol-gen} is just the condition of the deformations of Gerstenhaber for Lie algebras.
We explicitly prove \eqref{sol-gen} for $i=2$ the two remaining cases being similar.
If one checks only the terms in $t^2$, only the terms $\p + t \psi^{(1)} + t^2 \psi^{(2)}$ will matter. Inserting
\beqa
\p_{t\, 1}=\p_1 + t \psi^{(1)}_1 + t^2 \psi^{(2)}_1\nonumber\\
\p_{t\, 2}=\p_2 + t \psi^{(1)}_2 + t^2 \psi^{(2)}_2\nonumber\\
\p_{t\, 3}=\p_3 + t \psi^{(1)}_3 + t^2 \psi^{(2)}_3
\eeqa
\noi
in \eqref{defo-Jacobi}, the coefficient  of degree $1$ leads to
\beqa
\label{J2-t}
&&\ \ \ \p_2 (\psi_1^{(1)} (X_1, X_2), Y)+ \psi_2^{(1)} (\p_1 (X_1,X_2), Y) + \p_2 (\psi_2^{(1)} (X_2,Y),X_1)\nonumber \\
&&  +\ \psi_2^{(1)} (\p_2 (X_2,Y),X_1)+ \p_2 (\psi_2^{(1)} (Y, X_1), X_2) + \psi_2^{(1)} (\p_2 (Y, X_1), X_2)=0
\eeqa
\noi
and the coefficient of degree $2$ gives
\beqa
\label{J2-t^2}
&&\ \ \ \p_2 (\psi_1^{(2)} (X_1,X_2), Y) + \psi_2^{(2)} (\p_1 (X_1,X_2),Y)+\psi_2^{(1)} (\psi_1^{(1)} (X_1, X_2), Y) \nonumber \\
&&  +\ \p_2 (\psi_2^{(2)} (X_2,Y),X_1)+ \psi_2^{(2)} (\p_2 (X_2, Y), X_1)+\psi_2^{(1)} (\psi_2^{(1)} (X_2, Y), X_1) \nonumber\\
&&  +\ \p_2 (\psi_2^{(2)} (Y, X_1),X_2) + \psi_2^{(2)} (\p_2 (Y, X_1),X_2)+ \psi_2^{(1)} (\psi_2^{(1)}(Y, X_1),X_2)=0.
\eeqa
\noi
Then $\sum\limits_{p+q=1} \psi^{(p)} \circ_2 \psi^{(r)}=0$ and $\sum\limits_{p+q=2} \psi^{(p)} \circ_2 \psi^{(r)}=0$. 
Similarly, one proves \eqref{sol-gen} for any $r\in \NN^*$. QED

\label{defo-infi}

\begin{definition}
An infinitesimal deformation of $\p$ is a deformation
$\p_t$ of the form
$$\p_t=\p+t\psi^{(1)}.$$
\end{definition}

Let $\p_t=(\p_1+t\psi_1^{(1)}, \p_2+t\psi_2^{(1)}, \p_3+t\psi_3^{(1)})$. Identities 
\eqref{cond-defo} for the coefficient of $t$ lead to
\beqa
\label{defo-t}
&&  \p_1 (\psi_1^{(1)} (X_1, X_2), X_3)+\psi_1^{(1)} (\p_1 (X_1, X_2), X_3) +\p_1 
(\psi_1^{(1)} (X_3, X_1), X_2) \nonumber \\
& +& \psi_1^{(1)} (\p_1 (X_3, X_1), X_2) + \p_1 (\psi_1^{(1)} (X_2, X_3), X_1) +\psi_1^{(1)} (\p_1 (X_2, X_3), X_1)=0,\nonumber \\
\nonumber \\
&&\p_2 (\psi_1^{(1)} (X_1, X_2), Y)+ \psi_2^{(1)} (\p_1 (X_1,X_2), Y)  +\p_2 (\psi_2^{(1)} 
(X_2,Y),X_1)\nonumber \\
 & +& \psi_2^{(1)} (\p_2 (X_2,Y),X_1)+ \p_2 (\psi_2^{(1)} (Y, X_1), X_2) + \psi_2^{(1)} 
(\p_2 (Y, X_1), X_2)=0, \nonumber \\
 \\
&&\p_1 (X, \psi_3^{(1)} (Y_1, Y_2, Y_3))+\psi_1 ^{(1)}(X, \p_3 (Y_1, Y_2, Y_3)) 
  -\p_3 (\psi_2 ^{(1)}(X, Y_1), Y_2, Y_3) \nonumber \\
&-& \psi_3^{(1)} (\p_2 (X, Y_1), Y_2, Y_3) 
- \p_3 (Y_1, \psi_2^{(1)} (X,Y_2),Y_3)
 -\psi_3^{(1)} (Y_1, \p_2 (X,Y_2),Y_3)\nonumber \\
&-&\p_3 (Y_1, Y_2, \psi_2 ^{(1)}(X, Y_3)) -\psi_3^{(1)} (Y_1, Y_2, \psi_2^{(1)} (X, Y_3)) = 0,\nonumber\\
\nonumber\\
&&\p_2 (Y_1, \psi_3 ^{(1)}(Y_2, Y_3,Y_4)) + \psi_2 ^{(1)}(Y_1, \p_3 (Y_2, Y_3,Y_4))
+\p_2 (Y_2, \psi_3^{(1)} (Y_1, Y_3,Y_4)) \nonumber \\
&+& \psi_2^{(1)} (Y_2, \p_3 (Y_1, Y_3,Y_4)) 
+ \p_2 (Y_3, \psi_3^{(1)}(Y_1, Y_2,Y_4)) + \psi_2^{(1)} (Y_3, \p_3 (Y_1, Y_2,Y_4))\nonumber \\
&+&\p_2 (Y_4, \psi_3^{(1)} (Y_1, Y_2,Y_3)) + \psi_2 ^{(1)}(Y_4, \p_3 (Y_1, Y_2,Y_3))=0. \nonumber 
\eeqa
\noi
Using \eqref{def-criptic} these equations write
\beqa
\label{defo-t-criptic}
\p\circ_i \psi +\psi \circ_i \p =0, \mbox{ with }i=1,\dots,4, 
\eeqa
\noi
which is just equation \eqref{sol-gen} for $r=1$.

Furthermore, the coefficient of  $t^2$ obtained from \eqref{cond-defo} gives
\beqa
\label{defo-t2}
&&\psi_1^{(1)} (\psi_1^{(1)} (X_1, X_2),X_3)+\psi_1^{(1)} (\psi_1^{(1)} (X_3, X_1),X_2)+
\psi_1^{(1)} (\psi_1^{(1)} (X_2, X_3),X_1)=0,\nonumber\\
\nonumber\\
&&\psi_2^{(1)} (\psi_1^{(1)} (X_1, X_2), Y)+\psi_2^{(1)} (\psi_2^{(1)} (X_2, Y), X_1)+
\psi_2^{(1)} (\psi_2^{(1)} (Y,X_1), X_2)=0,\nonumber\\
\nonumber\\
&&\psi_1^{(1)} (X, \psi_3^{(1)} (Y_1, Y_2, Y_3))-\psi_3^{(1)}(\psi_2^{(1)} (X, Y_1),Y_2,Y_3) \\
&-& \psi_3^{(1)} (Y_1,\psi_2^{(1)} (X, Y_2), Y_3) -\psi_3^{(1)} (Y_1, Y_2, \psi_2^{(1)} 
(X,Y_3))=0,\nonumber\\
\nonumber\\
&&\psi_2^{(1)} (Y_1, \psi_3^{(1)} (Y_2, Y_3,Y_4)) + \psi_2^{(1)} (Y_2, \psi_3^{(1)} 
(Y_1, Y_2,Y_4))\nonumber\\
&+&\psi_2^{(1)} (Y_3, \psi_3^{(1)} (Y_1, Y_2,Y_4))+\psi_2^{(1)} (Y_4, \psi_3^{(1)} 
(Y_1, Y_2,Y_3))=0,\nonumber
\eeqa
\noi
which writes
\beqa
\psi^{(1)}\circ_i \psi^{(1)} = 0, \mbox{ with }i=1,\dots,4.
\eeqa

\begin{definition}
Denote by
$$Z(A)=\{ (\psi_1,\psi_2 ,\psi_3 ): A\to \g \},$$
where $\psi_i$ ($i=1,2,3$) satisfy \eqref{defo-t} and \eqref{defo-t2}. 
The vector space $Z(A)$ is called the infinitesimal deformation space of $A$.
\end{definition}

\subsection{Isomorphic deformations}

\begin{proposition}
Let $(\g=\g_0 \oplus \g_1,\p)\in {\cal F}_{m,n}$ be an elementary Lie algebra of order $3$. 
We consider a formal change
of basis given by
$\mathrm{Id}+tf_0\in GL(\g_0\otimes \mathbb C[[t]])$, $ \mathrm{Id} +tf_1\in GL(\g_1\otimes \mathbb C [[t]])$.  
The isomorphic multiplication $\p_t$ writes as the deformation
$$ \p_t = \p + t \psi + {\cal O} (t^2), $$ 
where $\psi = (\psi_1,\psi_2 ,\psi_3)$ is given by
\beqa
\label{deltele}
&&\psi_1(X_1, X_2)=\p_1 (f_0(X_1), X_2)+\p_1 (X_1,f_0 (X_2))-f_0 (\p_1 (X_1, X_2)), \nonumber \\
&&\psi_2(X, Y)=\p_2 (f_0(X), Y)+\p_2 (X,f_1 (Y))-f_1 (\p_2 (X,Y)), \nonumber \\
&&\psi_3(Y_1, Y_2, Y_3)=\p_3 (f_1(Y_1), Y_2, Y_3)+\p_3 (Y_1,f_1 (Y_2), Y_3)+\p_3 (Y_1, Y_2, f_0 (Y_3))-f_0 (\p_3 (Y_1, Y_2, Y_3)).\nonumber \\
\eeqa
\end{proposition}
{\it Proof.} 
We put 
\beqa
\label{noua-baza}
&&\tilde X_i =  ({\rm Id}+tf_0) (X_i)= X_i + t f_0 (X_i),\nonumber\\
&&\tilde Y_j =  ({\rm Id}+tf_1) (Y_j)= Y_j + t f_1 (Y_j),
\eeqa

and we have
\beqa
\p_1 (\tilde X_1, \tilde X_2)&=&h_0^{-1} \p_1 (h_0 (X_1), h_0 (X_2)), \nonumber \\
\p_2 (\tilde X_1, \tilde Y_2)&=&h_1^{-1} \p_2 (h_0 (X_1), h_1 (Y_2)), \nonumber \\
\p_3 (\tilde Y_1, \tilde Y_2, \tilde Y_3) &=& h_0^{-1} \p_3 ( g_1 (Y_1), g_1 (Y_2), g_1 (Y_3)).
\eeqa
\noi
This can be written as
\beqa
\p_1 (\tilde X_1, \tilde X_2)&=&\p_1 ( X_1, X_2)+t\psi_1(X_1, X_2) + {\cal O}(t^2), \nonumber \\
\p_2 (\tilde X, \tilde Y)&=&\p_2 ( X, Y)+t \psi_2 (X,Y) + {\cal O}(t^2), \nonumber \\
\p_3 (\tilde Y_1, \tilde Y_2, \tilde Y_3)&=&\p_3 ( Y_1, Y_2, Y_3)+t \psi_3 (Y_1, Y_2, Y_3) + {\cal O}(t^2),
\eeqa
\noi
where, by a tedious but straightforward calculation, one has \eqref{deltele}. QED

\begin{definition}
An elementary Lie algebra of order $3$ $\g=\g_0\oplus\g_1$ is called rigid if all deformations of $\g$ are  isomorphic to $\g$.
\end{definition}

If $\g$ is rigid then $\g_0$ is a rigid Lie algebra and the representation $\g_1$ of $\g_0$ is also rigid.

\medskip

As an example of rigid Lie algebra of order $3$ one has $\s \oplus {\rm ad}\, \s$ with $\alpha=1$ 
(see Theorem \ref{sl2-irrep} for notations). Other examples are given in subsections 
\ref{ex-contractie} and \ref{another}. 
An example of non-rigid Lie algebra of order $3$ is also exhibited in subsection \ref{another}.
Finally, note that some rigidity properties of representations of $\s$ can be found in \cite{folly}.

\medskip

\noindent {\bf Remark.} Usually, rigidy is computed using cohomological methods with the Nijenhuis-Richardson
theorem. This implies that we are able to define cohomology of algebra of order $3$ with values in a $\g$-module.
For instance, it is easy to define $1$ and $2$ cochains and the corresponding cocycles and coboundaries
space. For degree greater than 2, this is an open problem. For degree $2$, we can quickly summarize the construction. 

One denotes  $\psi$ (see Proposition 6.5) by $\delta_\p f$:
\beqa
\psi_1 (X_1, X_2)&=&(\delta_{\p_1} f)(X_1, X_2),\nonumber\\
\psi_2 (X_1,Y_2)&=&(\delta_{\p_2} f)(X_1, Y_2),\nonumber\\
\psi_3 (Y_1, Y_2, Y_3)&=&(\delta_{\p_3} f)(Y_1, Y_2, Y_3)
\eeqa
for any $X_i\in\g_0$ and $Y_j\in \g_1$. Let $Z(A)=Z^2(A)$ be the space of infinitesimal deformations (see Definition
6.3). Let $B^2(A)$ be the subspace of $Z^2(A)$ defined by
$$ B^2(A)=\{ \psi \in Z^2(A): \ \psi = \delta_\p f \}, $$
 we obviously have
$B^2(A)\subset Z^2 (A)$.
The theorem of Nijenhuis Richardson is written in this frame :

\noindent {\it If $H^2= Z^2/B^2=\{0\}$, then \el $\g$ is rigid.}

\subsection{Some deformations on the Poincar\'e algebra of order $3$}

Let us consider the algebra \eqref{lege-mica}. It leads to an explicit example of deformation. 
Let $\p=(\p_1,\p_2,\p_3)$ be 
the law defined by \eqref{lege-mica}. 
The deformation $\p_{t}=(\p_{t1},\p_{t\, 2},\p_{t\, 3})$ is given by
\beqa
\p_{t\, 1}(L_{mn}, L_{pq})&=& \p_1 (L_{mn}, L_{pq}),\nonumber\\
\p_{t\, 1}(L_{mn}, P_{p})&=& \p_1 (L_{mn}, P_{p}),\nonumber\\
\p_{t\, 1}(P_{m}, P_{p})&=& -t^2 L_{mp},\nonumber\\
\p_{t\, 2}(L_{mn}, V_{pq})&=& \p_2 (L_{mn}, V_{pq}),\nonumber\\
\p_{t\, 2}(L_{mn}, V_{p})&=& \p_2 (L_{mn}, V_{p}),\nonumber\\
\p_{t\, 2}(P_{m}, V_{pq})&=& t(\eta_{mp} V_q - \eta_{mq} V_p), \nonumber\\
\p_{t\, 2}(P_{m}, V_{p})&=& t V_{mp}, \nonumber\\
\p_{t\, 3} (V_{mn},V_{pq}, V_{rs})&=&t((\eta_{mp}\eta_{nq}-\eta_{mq}\eta_{np}) L_{rs}+ 
(\eta_{mr}\eta_{ns}-\eta_{ms})L_{pq}+ (\eta_{pr}\eta_{qs}-\eta_{ps}\eta_{qr})L_{mn}),\nonumber\\
\p_{t\, 3}(V_{mn},V_{pq}, V_{r})&=&\p_3 (V_{mn},V_{pq}, V_{r}),\nonumber\\
\p_{t\, 3}(J_{mn},P_p, P_{r})&=&t\, \eta_{pr} L_{mn},\nonumber\\
\p_{t\, 3}(V_{m},P_{p}, P_{r})&=&\p_3 (V_m, V_p, V_r).\nonumber
\eeqa
\noi
and we obtain \eqref{p-so23}.

\bigskip

{\bf Acknowledgment:}  We would like to thank M. J. Slupinski for help in proving Lemma \ref{rep}.

\end{document}